\documentclass{emulateapj}

\usepackage{amssymb}
\usepackage{amsmath}
\usepackage{natbib}
\usepackage{txfonts}
\usepackage[colorlinks=true,citecolor=blue,linkcolor=blue]{hyperref}
\usepackage{microtype}
\usepackage{color}
\usepackage{graphicx}
\usepackage{epstopdf}
\usepackage{multirow}
\usepackage{tabularx}

\def\apjl{ApJL}
\def\apjs{ApJS}

\def\aap{A\&A}

\usepackage{soulutf8} % strikethrough com cor
\usepackage[usenames,dvipsnames]{xcolor}
\def\sgr{\mbox{SGR~1745-2900\,}}
\def\ssgr{\mbox{SGR~1745-2900}}
\def\psr{\mbox{PSR J0030+0451}}

\shorttitle{Evidence for a multipolar magnetic field in SGR J1745--2900 from X-ray light-curve analysis}
\shortauthors{de Lima et al.}

\begin{document}

\title{Evidence for a multipolar magnetic field in SGR J1745--2900 from X-ray light-curve analysis}

\author{Rafael C. R. de Lima\altaffilmark{1,2}, Jaziel G. Coelho\altaffilmark{2,3,4}, Jonas P.~Pereira\altaffilmark{5,6,7}, Claudia V. Rodrigues\altaffilmark{4}, and Jorge A. Rueda\altaffilmark{2,8,9,10,11}}

%ORCID - CLAUDIA - 0000-0002-9459-043X

\altaffiltext{1}{Universidade do Estado de Santa Catarina, Av. Madre Benvenuta, 2007 Itacorubi, 88.035-901, Florian\'opolis, Brazil}

\altaffiltext{2}{ICRANet, Piazza della Repubblica 10, I--65122, Pescara, Italy}

\altaffiltext{3}{Departamento de F\'isica, Universidade Tecnol\'ogica Federal do Paran\'a, 85884-000 Medianeira, PR, Brazil}

\altaffiltext{4}{Divis\~ao de Astrof\'isica, Instituto Nacional de Pesquisas Espaciais, Avenida dos Astronautas 1758, 12227-010, S\~ao Jos\'e dos Campos, SP, Brazil}

\altaffiltext{5}{Universidade Federal do ABC, Centro de Ci\^encias Naturais e Humanas, Avenida dos Estados, 5001- Bang\'u, CEP 09210-170, Santo Andr\'e, SP, Brazil}

\altaffiltext{6}{Mathematical Sciences and STAG Research Centre,
University of Southampton, Southampton, SO17 1BJ, United Kingdom}

\altaffiltext{7}{Nicolaus Copernicus Astronomical Center, Polish Academy of Sciences, Bartycka 18, 00-716, Warsaw, Poland}

\altaffiltext{8}{ICRANet-Rio, Centro Brasileiro de Pesquisas F\'isicas, Rua Dr. Xavier Sigaud 150, 22290-180 Rio de Janeiro, Brazil}

\altaffiltext{9}{ICRANet-Ferrara, Dipartimento di Fisica e Scienze della Terra, Universit\`a degli Studi di Ferrara, Via Saragat 1, I--44122 Ferrara, Italy}

\altaffiltext{10}{Dipartimento di Fisica e Scienze della Terra, Universit\`a degli Studi di Ferrara, Via Saragat 1, I--44122 Ferrara, Italy}

\altaffiltext{11}{INAF, Istituto di Astrofisica e Planetologia Spaziali, Via Fosso del Cavaliere 100, 00133 Rome, Italy}

\begin{abstract}
SGR J1745--2900 was detected from its outburst activity in April 2013 and it was the first soft gamma repeater (SGR) detected near the center of the Galaxy (Sagittarius A$^*$). We use 3.5-year Chandra X-ray light-curve data to constrain some neutron star (NS) geometric parameters. We assume that the flux modulation comes from hot spots on the stellar surface. Our model includes the NS mass, radius, a maximum of three spots of any size, temperature and positions, and general relativistic effects. 
We find that the light-curve of SGR J1745--2900 could be described by either two or three hot spots. The ambiguity is due to the small amount of data, but our analysis suggests that one should not disregard the possibility of multi-spots (due to a multipolar magnetic field) in highly magnetized stars.
For the case of three hot spots, we find that they should be large and have angular semi-apertures ranging from $16$--$67$ degrees.
The large size found for the spots points to a magnetic field with a nontrivial poloidal and toroidal structure (in accordance with  magnetohydrodynamics investigations and NICER's recent findings for  \psr) and is consistent with the small characteristic age of the star. Finally, we also discuss possible constraints on the mass and radius of SGR J1745--2900 and briefly envisage possible scenarios accounting for the 3.5-year evolution of SGR J1745--290 hot spots.
\end{abstract}


\altaffiltext{}{rafael.lima@udesc.br,jazielcoelho@utfpr.edu.br,jpereira@camk.edu.pl \\}

\maketitle

\section{Introduction}
\label{intt}

Electromagnetic data-driven constraints to the mass and radius of NSs are very elusive. 
Radius measurements are mostly
based on the observation of thermal emission and comparisons with theoretical models.
The modeling, however, due to the complex and relativistic nature of NSs, suffers from a number of complications such as parameter degeneracy, the unknown NS equation of state (EOS), among other uncertainties, e.g. the distance to the object \citep[see, e.g.,][and references therein]{2016ApJ...820...28O,2016ARA&A..54..401O}.
Notwithstanding, currently operating and future observatories, such as the Neutron Star Interior Composition Explorer (NICER)~ \citep{2016SPIE.9905E..1HG}, the enhanced X-ray Timing and Polarimetry mission (eXTP) \citep{2019SCPMA..6229502Z}, and the Spectroscopic Time-Resolving Observatory for Broadband Energy X-rays (STROBE-X)~\citep{2018SPIE10699E..19R}, promise to greatly decrease the uncertainties of NS parameters. They are expected to provide
masses and radii of NSs with an accuracy of a few percent \citep[see][and references therein]{2018A&A...616A.105S}.
In particular, one of the most significant developments in the measurement of the dense matter EOS is going to come from the NICER detector~\citep[see][]{2016ApJ...832...92O}. 
The pulsed X-ray emission from hot spots on the surface of a rotating NS contains encoded information about its gravitational field and the properties of the spot emission pattern. NICER is using this approach to measure NS radii, based on the shape and amplitude of the pulsed emission observed from pulsar surface in multiple wavebands. The data accuracy allows for precise comparison between measurements and models of NSs~\citep{2018A&A...616A.105S}, and will significantly improve our understanding of the physics of superdense matter in the universe.
Indeed, NICER's X-ray data from \psr \, has recently led to the first precise measurements (below $10\%$ uncertainty) of the radius and mass of a pulsar \citep[see][]{Bilous_2019,Riley_2019,Raaijmakers_2019,Miller_2019,Bogdanov_2019,Bogdanov_2019_,Guillot_2019}. Besides, it has also allowed for the first map of the hot spots on the surface of a star. It provided the locations, shapes, sizes, and temperatures of the heated regions, which should give precise details of the magnetic field of a neutron star. In this regard, it has already been found that the hot spots are far from antipodal, meaning that the magnetic field structure of a compact star is much more complex than previously thought.

In order to constrain uncertainties up to a few percent, stellar rotation should be large ($>100$ Hz), time resolution should be small ($\lesssim 10\mu$s), and the
number of photons should be large (at least $\sim 10^6$) \citep{2019arXiv190407012W}. However, it is still possible to obtain interesting constraints on the properties of slowly rotating neutron stars, such as the Soft Gamma Repeaters (SGRs) and the Anomalous X-ray Pulsars (AXPs).

\sgr~was the first Soft Gamma Repeater detected near the Milky Way center, Sagittarius A$^{*}$~\citep{Kennea_2013,2013ApJ...770L..23M}, and it is at distance of $8.3$~kpc. It has a rotational period $P = 3.76$~s and a changing spindown rate since the 2013 outburst. From its latest update, it is $\dot{P}\sim 3\times 10^{-11}$~s/s \citep{2017MNRAS.471.1819C}. It is characterized by a X-ray luminosity $L_X \approx 10^{32}$--$10^{36}$~erg~s$^{-1}$. Owing to the flaring/outburst activity ($10^{38}$--$ 10^{45}$~erg), \sgr has been classified within the SGR and AXP class \citep[see, e.g.,][]{Olausen_2014}. For a comprehensive review on observations of \ssgr, even the long-term ones, see \cite{2015MNRAS.449.2685C,2017MNRAS.471.1819C}. For a systematic study of pulsed fractions of magnetars in quiescent state, including \sgr, see \cite{2019MNRAS.485.4274H}.

In this paper, we apply the approach of \citet{2013ApJ...768..147T} for the emission of a NS with hot spots to two X-ray light-curves of \sgr in different epochs. We use Genetic Algorithm techniques to constrain the mass and radius of \sgr with a minimum set of assumptions.  
This paper is organized as follows. In Sec. \ref{sec_Flux}, we present the aspects of the model used for obtaining light-curves from NS surfaces with hot spots. Section \ref{GA} explains the genetic algorithm techniques we use for fits of the \sgr light-curves and how to obtain the NS parameters. In Secs. \ref{Results} and \ref{discussion} we present our results and discuss them.

%%%%%%%%%%%%%%%%%%%%%%%%%%%%%%%%%%%%%%%%%%%%%%%%%%%%%%%%%%%%%%%%%%%%
%%%%%%%%%%%%%%%%%%%%%%%%%%%%%%%%%%%%%%%%%%%%%%%%%%%%%%%%%%%%%%%%%%%%
\section{Pulsed Profile model}
\label{sec_Flux}
%%%%%%%%%%%%%%%%%%%%%%%%%%%%%%%%%%%%%%%%%%%%%%%%%%%%%%%%%%%%%%%%%%%%
%%%%%%%%%%%%%%%%%%%%%%%%%%%%%%%%%%%%%%%%%%%%%%%%%%%%%%%%%%%%%%%%%%%%

\begin{figure}
\centering
\includegraphics[width=0.6\hsize,clip]{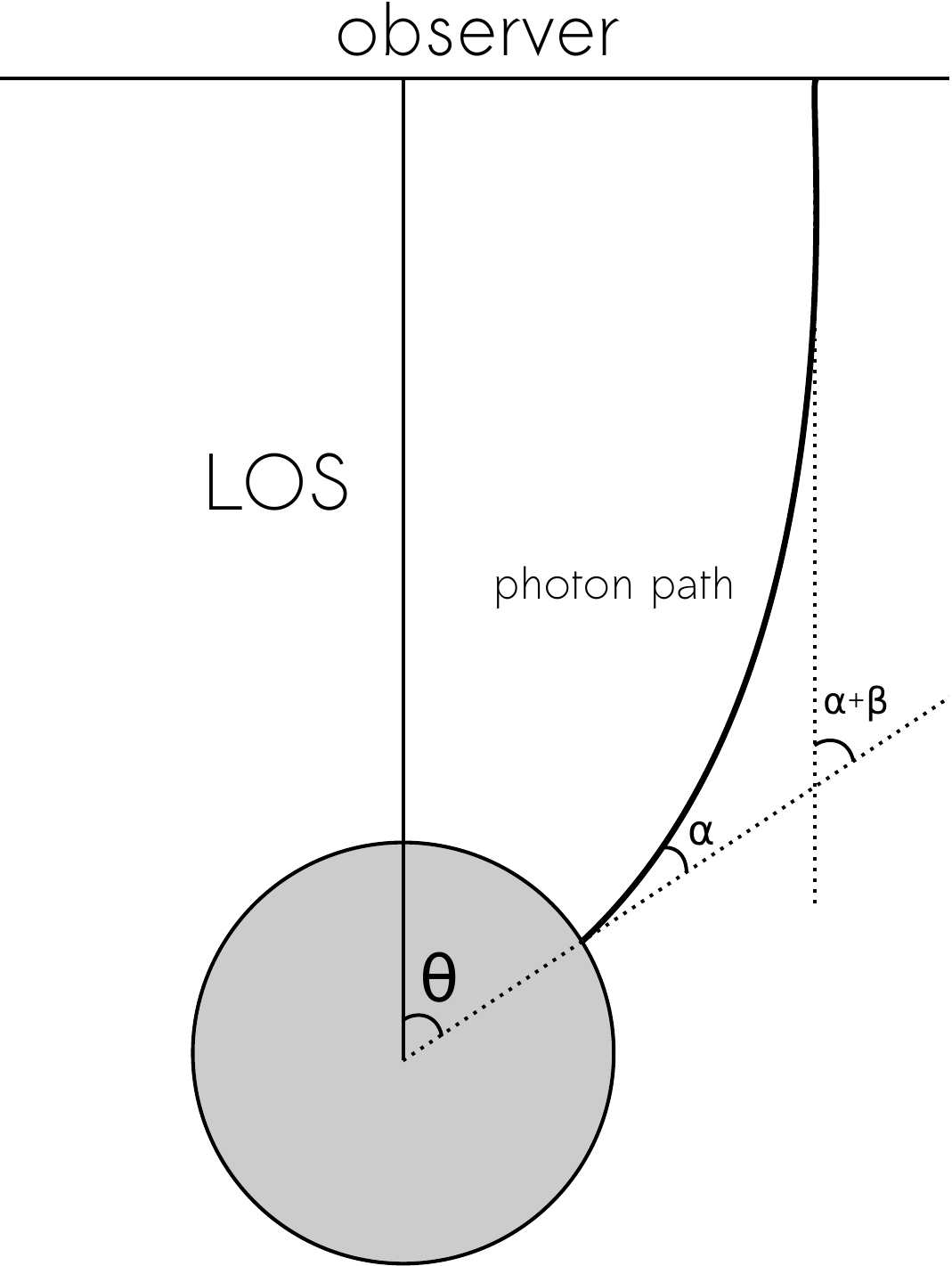}
\caption{Illustration of the model geometry showing the photon trajectory and the angles $\theta$, $\alpha$, and $\beta$. }
\label{photonpath}
\end{figure}

Here we show how the theoretical pulsed profiles are calculated for a NS with thermal spots on its surface.
We follow the procedure of \citet{2013ApJ...768..147T} to calculate the observed flux, which allows us to treat circular spots having arbitrary size and location on the stellar surface. The mass and radius of the star are denoted by $M$ and $R$, respectively, and the spacetime outside the star is described by the Schwarzschild metric, i.e. we neglect rotational effects. This is an accurate approximation for SGR J1754--2900 given its slow rotational period of $3.76$~s \citep[clearly contrasting with millisecond pulsars, see e.g.][]{2015ApJ...799...23B,2015PhRvD..92b3007C,2017A&A...599A..87C}. Let $(r,\theta,\phi)$ be a spherical coordinate system with the origin at the stellar center and the polar axis along the line of sight (LOS)~(see Fig.~\ref{photonpath}).
We consider an observer at $r\rightarrow \infty$ and a photon that arises from the stellar surface at $dS=R^2 \sin\theta d\theta d\phi$, making an angle $\alpha$ with the local normal to the surface ($ 0 \leq \alpha \leq \pi/2$). The photon path is then bended by an additional angle $\beta$ owing to the spacetime curvature, and the effective emission angle as seen by the observer is  $\psi=\alpha+\beta$ (see Fig.~\ref{photonpath}). The geometry is symmetric relative to $\phi$.  \citet{2002ApJ...566L..85B} has shown that the following simple approximate formula can be used to relate the emission angle $\alpha$ to the angle $\theta$:
\begin{equation}\label{Beloborodov_approximation}
    1 -\cos\alpha = (1-\cos\theta)\left(1-\frac{R_s}{R}\right),
\end{equation}
where $R_s=2GM/c^2$ is the Schwarzschild radius and $G$ denotes the gravitational constant.  We note that Eq.~(\ref{Beloborodov_approximation}) is a very good approximation for $R>3R_s$ since it typically leads to very small errors ($\lesssim 1\%$). For the range of masses and corresponding radii of interest here, errors would be up to a few percent.

We assume that the spot emission follows a local Planck spectrum and that the observed flux comes mainly from hot spots.
The intensity $B_\nu(T)$ is given by a blackbody with temperature $T$, where $\nu$ is the photon frequency. The flux is proportional to the visible area of the emitting region ($S_V$) plus a relativistic correction, and it is given by \citep{2002ApJ...566L..85B,2013ApJ...768..147T}
\begin{align}\label{FluxTurolla}
    F_\nu &= \left(1-\frac{R_s}{R}\right)^2 B_\nu(T) \int_{S_V} \cos\alpha \frac{d\cos\alpha}{d(\cos\theta)}ds \nonumber \\
    &=\left(1-\frac{R_s}{R}\right)^2 B_\nu (T)(I_p + I_s),
\end{align}
where
\begin{equation}
 I_p = \int_{S_V} \cos\theta \sin\theta d\theta d\phi,\quad I_s = \int_{S_V} \sin\theta d\theta d\phi .
\end{equation}
In polar coordinates, the circular hot spot has its center at $\theta_0$ and a semi-aperture $\theta_c$. The spot is bounded by the function $\phi_b(\theta)$, where $0\leq \phi_b \leq \pi$, and since we must consider just the visible part of the star, the spot must be also limited by a constant $\theta_F$. It is defined by
\begin{equation}\label{thetaF}
    \theta_{F}=\arccos{\left( 1- \frac{c^2 R}{2 G M} \right)^{-1}}.
\end{equation}
For a given bending angle $\beta$, $\theta_F$ occurs for the maximum emission $\alpha$, i.e. $\alpha=\pi/2$. In Newtonian gravity, where $\beta=0$, the maximum visible angle is $\theta_F=\pi/2$, meaning that half of the stellar surface is visible. However, for a relativistic star $\theta_F>\pi/2$. 
Then
\begin{align}
    & I_p = 2 \int_{\theta_{min}}^{\theta_{max}} \cos\theta \sin\theta \phi_b(\theta) d\theta, \nonumber \\
    & I_s = 2 \int_{\theta_{min}}^{\theta_{max}} \sin\theta \phi_b(\theta) d\theta,
\end{align}
where $\theta_{min}$, $\theta_{max}$ are the limiting values, to be determined to the spot considered.
\citet{2013ApJ...768..147T} show how to solve these integrals and how to carefully treat the limiting angles. 
Finally, the flux given by Eq. \eqref{FluxTurolla} can be written as \citep{2013ApJ...768..147T} 
\begin{equation}\label{Flux}
    F_\nu = \left(1-\frac{R_s}{R}\right)^2 \frac{B_\nu(T)}{D^2} A_{\rm eff}(\theta_c,\theta_0) \ ,
\end{equation}
where $D$ is the distance to the source, and it corrects the flux for an observer on Earth, and $A_{\rm eff}$ is the effective area, given by
\begin{equation}\label{Aeff}
    A_{\rm eff}(\theta_c,\theta_0) = R^2 \left[ \frac{R_s}{R}I_s + \left(1-\frac{R_s}{R}\right)I_p\right] \ .
\end{equation}
 The total flux produced by $N_\sigma$ spots, where the $\sigma$-th spot has a semi-aperture $\theta_{c\sigma}$ and a temperature $T_\sigma$, can be calculated by adding up each contribution, and so we have
\begin{eqnarray}\label{TotalFlux}
    F_\nu^{TOT} &=&  \left(1-\frac{R_s}{R}\right)^2 \sum_\sigma  \ \frac{B_\nu(T_\sigma)}{D^2} A_{\rm eff}(\theta_{c\sigma},\theta_{0\sigma}) \ .
\end{eqnarray}
Besides, the pulse profile in a given energy band $[\nu_1,\nu_2]$ for a given spot $\sigma$ is

\begin{equation}
    F_\sigma(\nu_1,\nu_2)=\left(1-\frac{R_s}{R} \right)^2 A_{\rm eff}(\theta_{c\sigma},\theta_{0\sigma}) \int_{\nu_1}^{\nu_2} \frac{B_\nu(T_\sigma)}{D^2}d\nu \ .
\end{equation}
Therefore, one can rewrite Eq.~(\ref{TotalFlux}) for a given energy band, and it becomes
\begin{eqnarray}\label{TotalFluxBand}
    F^{TOT}  = \sum_\sigma F_\sigma(\nu_1,\nu_2)  \ .
\end{eqnarray}

We define by $\hat{\mathbf r}$ the unit vector parallel to the rotation axis of the star, whose angular velocity is $\Omega=2\pi/P$. It is also useful to introduce $i$, the angle between the LOS (unit vector $\hat{\mathbf l}$) and the rotation axis, and $j$, the angle between the polar cap axis (unit vector $\hat{\mathbf c}$) and the rotation axis ($\cos i=\hat{\mathbf r}\cdot\hat{\mathbf l}$ and $\cos j=\hat{\mathbf r}\cdot\hat{\mathbf c}$).

When the total flux, Eq.~(\ref{TotalFluxBand}), is calculated for a given configuration $(i,j)$ for a time interval $(0 - P)$, the typical result is a pulsed flux with a maximum ($F_{\rm max}$) and a minimum flux ($F_{\rm min}$). We shall use the normalized version of Eq.~(\ref{TotalFluxBand}), given by

\begin{eqnarray}\label{TotalFluxBandNorm}
    \bar{F}^{TOT}  = \frac{1}{\bar{N}} F^{TOT} \ ,
\end{eqnarray}
\noindent
where $\bar{N}=(F_{max}+F_{min})/2$. 
This normalization makes our model independent of the source distance, avoiding uncertainties linked to its precise determination. As \sgr is located near the Galactic Center, its emission is heavily absorbed by the interstellar medium (ISM). However, we have verified that the ISM absorption can be neglected when using this normalization. 

We also define  the pulsed fraction as
\begin{equation}\label{PF}
{\rm PF}=\frac{F_{\rm max}-F_{\rm min}}{F_{\rm max}+F_{\rm min}}.
\end{equation}

We have considered two main physical scenarios. (i) Two-spot configuration: the spots can have any size and temperature, but their centers are diametrically opposed (as the poles of a dipolar magnetic field). So, in this case, the spots are called polar caps and we can define a polar cap axis. (ii) Three-spot configuration: two-spot configuration plus a third spot of any size, location, and temperature.

As the star rotates, the polar coordinate of the spot's center, $\theta_0$, changes. Let $\gamma(t)=\Omega t$ be the star's rotational phase. Thus, from a geometrical reasoning we have that
\begin{equation}\label{thetat}
    \cos\theta_0(t)=\cos i\cos j-\sin i\sin j\cos\gamma(t) \ ,
\end{equation}
where we have taken that $i$ and $j$ do not change with time.

%%%%%%%%%%%%%%%%%%%%%%%%%%%%%%%%%%%%%%%%%%%%%%%%%%%%%%%%%%%%%%%
\section{Genetic Algorithms}
\label{GA}
A Genetic Algorithm (GA) is a type of programming technique inspired in the modern understanding of natural selection, i.e. the best genetic code is the one whose phenotype manages to survive all natural vicissitudes. In our work, the {\it{chromosome}} is given by the set of all free parameters used to generate a theoretical pulse profile. In GA, the individual parameters of a chromosome are called {\it genes}. In our case, the mass and radius of the star ($M$ and $R$) and the angles $i$ and $j$ are examples of genes. The entire set of genes is given in Table~(\ref{chromo}).

The desired phenotype is given by the observed pulse profile, and a chromosome fitness is calculated from it. A typical GA procedure comprises six steps:
\begin{itemize}
    \item[1)] Initialization: generation of a population of solutions (i.e. the chromosomes);
    \item[2)] Phenotype evaluation - calculation of each model solution fitness;
    \item[3)] Selection of the best solutions;
    \item[4)] Reproduction - the genes of the best solutions are recombined;
    \item[5)] Mutation - genes can be randomly selected and changed;
    \item[6)] Population replacement.
\end{itemize}

Every iteration from step 2 to 6 is called a {\it generation}. In order to handle the genetic evolution and gene operations, we use the python library {\bf Pyevolve} \footnote{ {\UrlFont{http://pyevolve.sourceforge.net/}}}, maintained by Christian S. Perone and modified by us.

\subsection{Goodness-of-fit calculation}

The goodness-of-fit (GoF) of a given solution is  calculated by the square of the difference between the model and the observed data. This is summed over the period of the pulsed profile, i.e., \\
\begin{equation}\label{Phi}
    \text{GoF} = \sum_{k} \left[ \bar{F}^{TOT}_k-\bar{F}^{OBS}_k \right] ^2 \ , 
\end{equation}
\noindent
where $\bar{F}^{TOT}_k$ is given by Eq.~(\ref{TotalFluxBandNorm}). Note that the summation is discrete because of the data nature, but the temporal change in  $\bar{F}^{TOT}_k$ is controlled by Eq.(\ref{thetat}) over the star's period. $\bar{F}^{OBS}_k$ is the normalized observed flux and $k = 1$--$N$, where $N$ is the number of observed points of the light-curve. The optimal case would be $\text{GoF}=0$. Therefore, the GA's goal is to minimize GoF. We note that the data uncertainty $\sigma$ of \sgr is a given constant for each dataset, and hence GoF and the standard $\chi^2$ ($\chi^2\equiv  \text{GoF}/\sigma^2$) carry the same statistical information. Since the definition given by Eq. \eqref{Phi} is better suited for numerical computations, we use it for our fits. However,
for statistical considerations we use $\chi^2$  in order to be closer to standard analyses.

\begin{table}
\centering
\begin{small}
\begin{tabular}{l c c}
\hline
\multicolumn{3}{c}{Chromossome}\\
	\hline
		 Gene & Definition & Range \\
          \hline
		 $M(M_{\odot})$ & Star's Mass & $1.0-2.0$ \\
		 $R$~(km) & Star's Radius & $8.9-13.7$  \\
		 $N_{\sigma}$ & Number of hot spots & $1-4$ \\		
		 $\theta_{c\sigma}$ & $\sigma$-th spot's  semi-aperture  & $2-180\degr$\\
		 $T_{\sigma}$~(keV) & $\sigma$-th spot's temperature & $0.0-0.9$ \\
		 $\theta_{\sigma}$ & $\sigma$-th spot's colatitude & $0-180\degr$   \\
		 $\phi_{\sigma}$ & $\sigma$-th spot's longitude & $0-360\degr$ \\
		 $i$& angle between the LOS  & $0-90\degr$\\
		 & and the rotation axis  & \\
		 $j$ & angle between the polar cap  & $0-90\degr$\\
		 & and the rotation axis  & \\
		 \hline
\end{tabular}
\end{small}
\caption{List of parameters and ranges used in our genetic algorithm to fit a light-curve. }\label{chromo}
\end{table}	
\section{Results}
\label{Results}
Our aim is to find the set of parameters~(see Table \ref{chromo}) that best fit the X-ray emission of SGR J1745--2900. We use the light-curve from two epochs: 2013 ({\it D13}) and 2016 ({\it D16})\, -- presented by
\citet{2017MNRAS.471.1819C}. We let the parameters evolve as laid out in Section~\ref{GA}, and this is done independently for each data set. The final criterion to accept the best solutions is that both {\it D13} and {\it D16} result in the same most likely radius and inclination angles $i$ and $j$, since these are expected to remain stable. For the determination of the mass and radius (based on the mean mass) ranges, global data analyses have been done, as explained below.

\begin{figure*}[h!]
\centering
\includegraphics[width=0.5\hsize,clip]{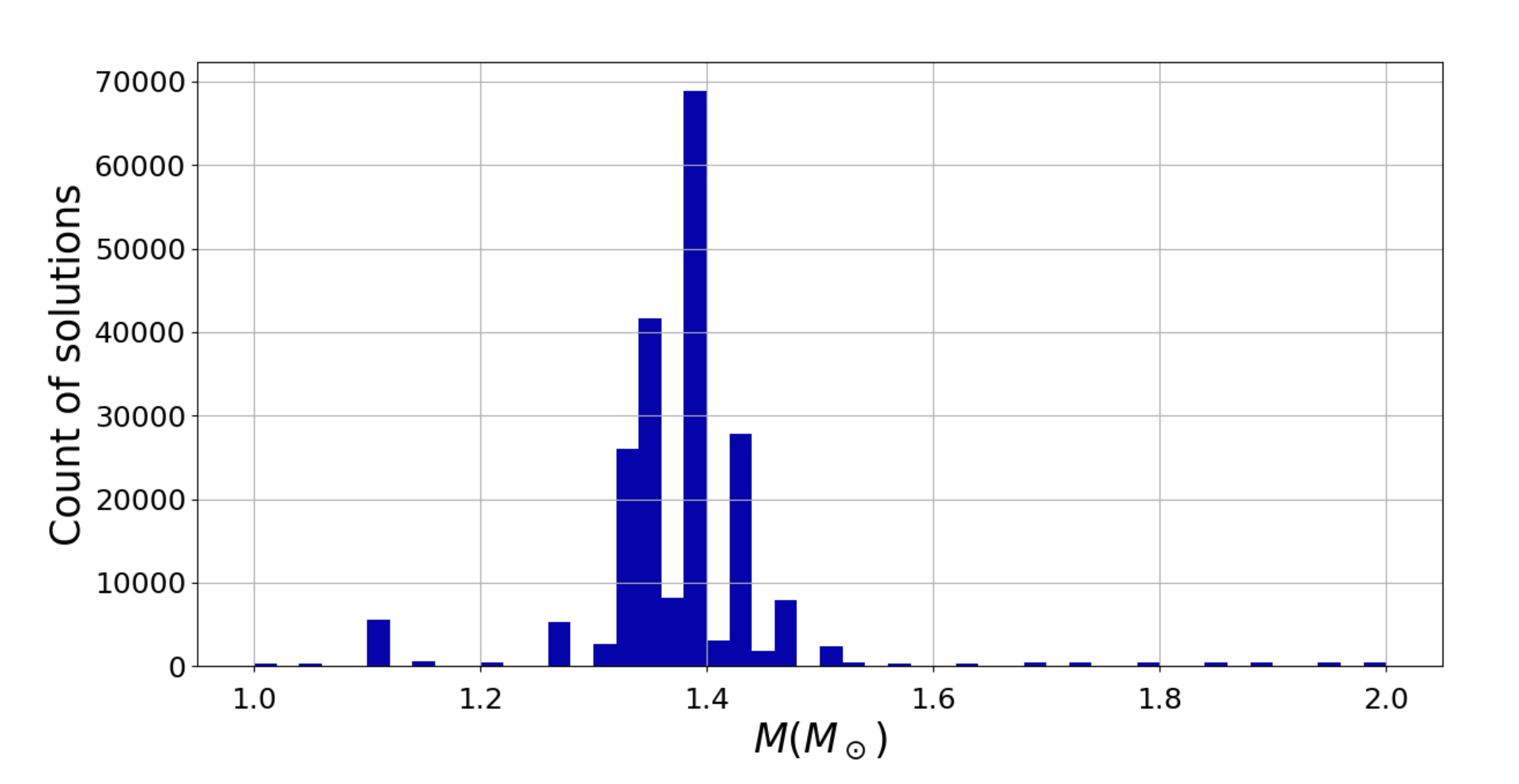}\includegraphics[width=0.5\hsize,clip]{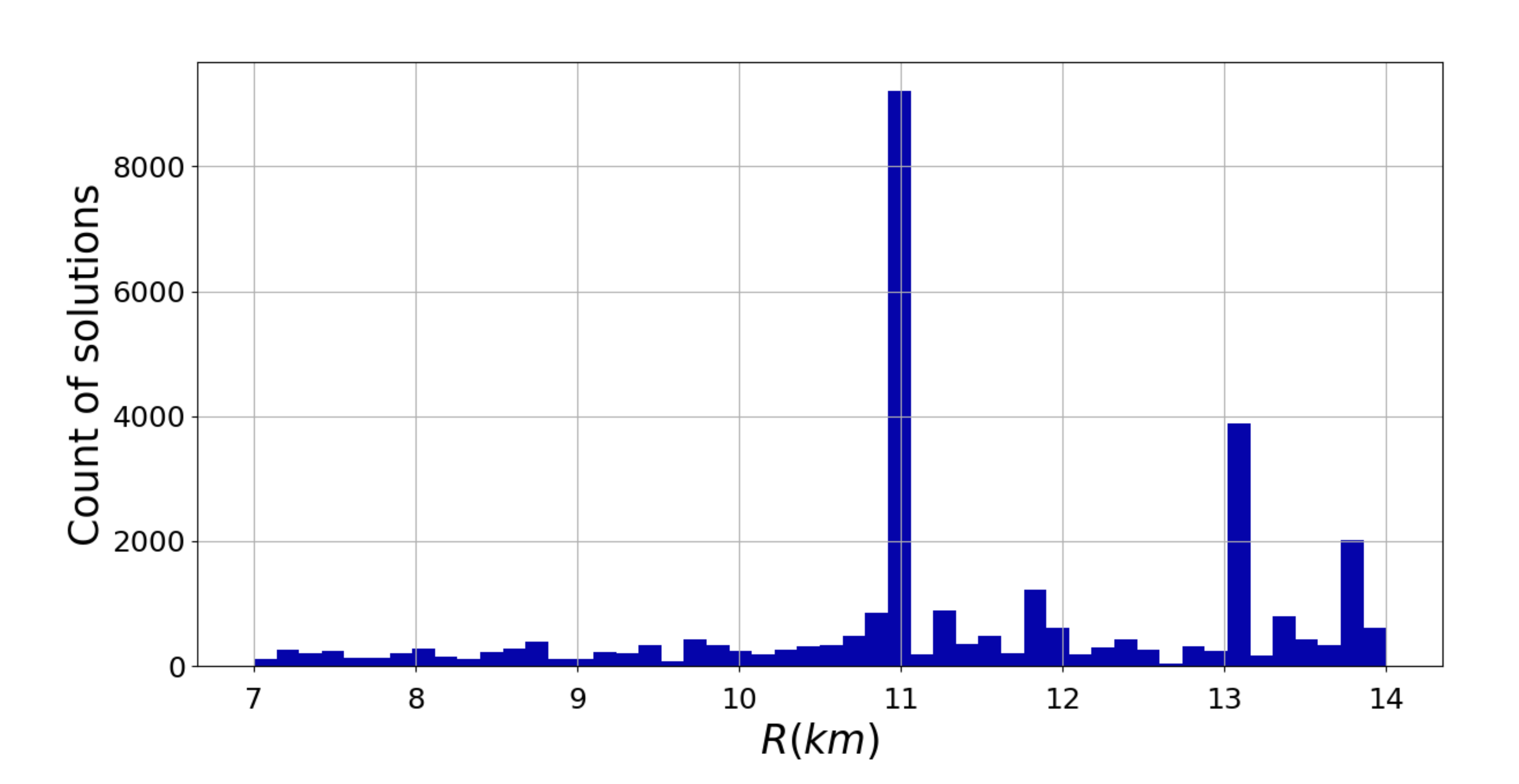}
\caption{Histograms of all generations of solutions for D13 and D16. Left panel: From a normal distribution fit of the count of solutions one learns that the mean mass is $1.4 M_\odot$ and the standard deviation is $0.1 M_\odot$. Right panel: The histogram shows the count of solutions of different radii for $M=1.4~M_{\odot}$ (the mean mass). Also from a normal distribution fit, the mean radius in this case is given by $R=10.9$ km and the standard deviation is $1.5$ km.}
\label{fig:HistMass861}
\end{figure*}

We have performed a ``zeroth run'' with all data points to find out which values of mass were the most likely to fit the \sgr~light-curve. This has been done in order to fix one parameter and expedite the convergence time of subsequent (more precise) analyses. Our results are summarized in Fig.~\ref{fig:HistMass861} where one has the histogram of {\it all generations of solutions} fitting \sgr~light-curves. 
There one sees that to one standard deviation, the majority of candidates have mass $1.4\pm 0.1$~M$_{\odot}$. 
Thus, we take the \sgr~mass as a fixed value in the subsequent fits and equal to the mean value of the normal distribution of Fig.~\ref{fig:HistMass861}, the canonical NS ($1.4M_\odot$). However, as the large radius scattering of the zeroth-run (when compared to the mass) already suggests ($R= 10.9\pm 1.5$ km), we do not take the radius of \sgr as a fixed parameter in our subsequent investigations. Further details in this regard are given in Sec. \ref{discussion}.

\begin{figure*}[h]
\centering
\includegraphics[width=1\hsize,clip]{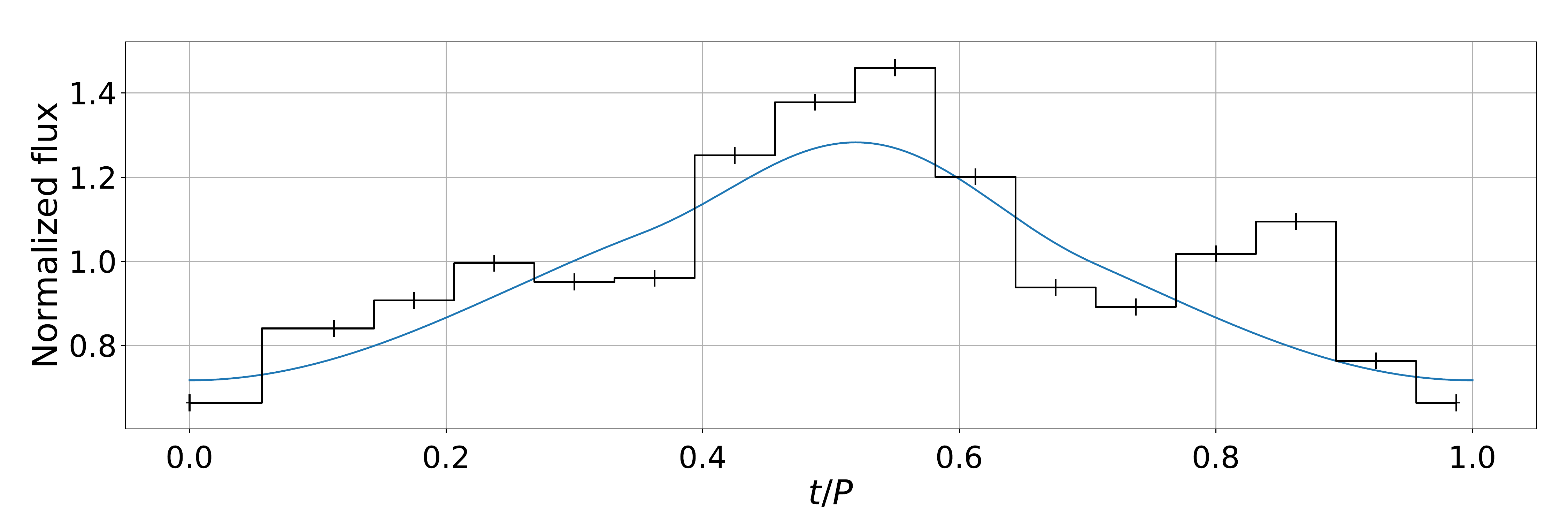}
\includegraphics[width=1\hsize,clip]{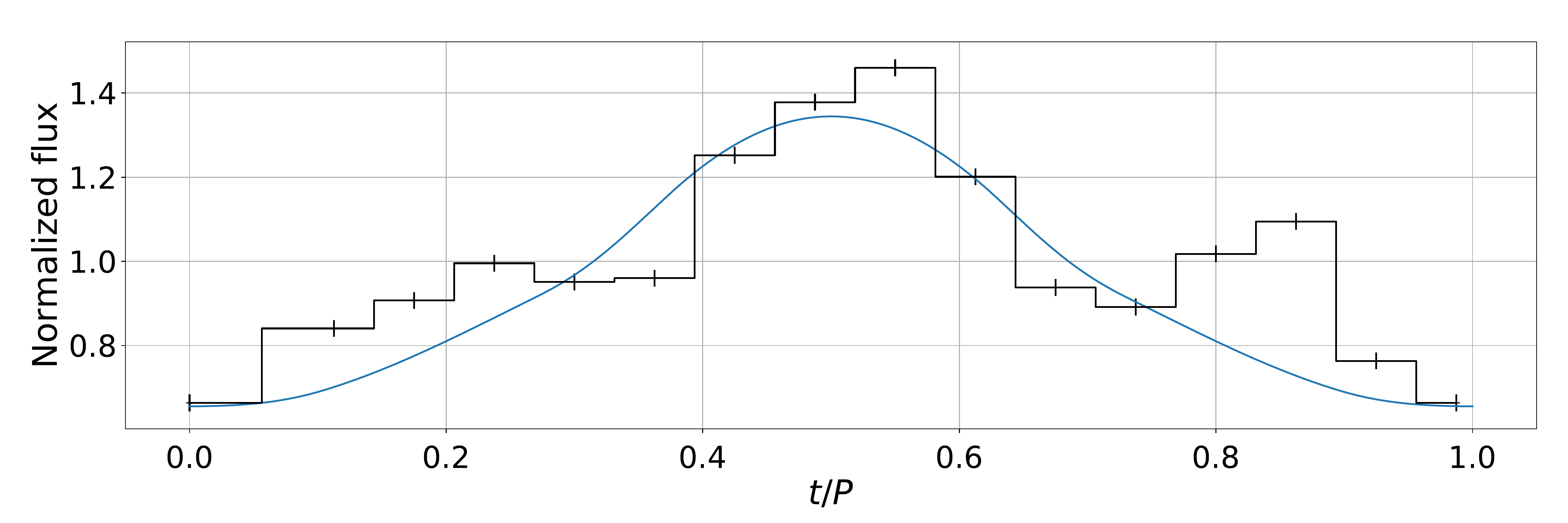}
\caption{Upper panel: D13's fitting for two spots. The mass is $1.4~M_{\odot}$ and the two spots are free. The radius found is $R=13.74~$km, and GoF = $0.22$. In this case, the number of degrees of freedom (DoF) is 5 and hence $\text{GoF}/\text{DoF}=0.044$. Bottom panel: D13's fitting for two antipodal spots. The mass is $1.4~M_{\odot}$, $R=13.4~$km, and GoF = $0.29$. Here, $\text{DoF}= 7$ and then $\text{GoF}/\text{DoF}=0.041$. The normalization factor used in the plots is $\bar{N}=(F_{max}+F_{min})/2$.}
\label{two-hot-spots}
\end{figure*}

As a first test, we have attempted to fit the light-curve with only one hot spot, but the fits were very poor and are not discussed here. So we explore two spots, either having free positions or being antipodal.
The two-spot fits can be seen in Fig. \ref{two-hot-spots} for the {\it D13} dataset, where
the GoF per degree of freedom for the fits are in the range 0.041 -- 0.044.
In order to contemplate another geometry, we added a third hot spot with a free position relative to the other two, chosen to be antipodal. This choice of spots acts like a correction (which can be large) to the dipolar model, and, as shown below, it results in better fits to the light-curves. 
A summary of the best-fit parameters for the {\it D13} and {\it D16} data sets in this case can be seen in Table ~\ref{best_chromo}.
Figure \ref{fig:gen837} shows the best fits for the {\it D13} and {\it D16} sets using three spots. One can see that three spots fit reasonably well the main features of both data sets. For the {\it D13} dataset we find that $\text{GoF}$ per degree of freedom is around 0.037, which is slightly better than the two spot fits. We discuss further the quality of the fits and some subtleties of the {\it D16} dataset in Secs. \ref{statistics} and \ref{discussion}.

Figure \ref{fig:star844} shows the hot spot positions on the stellar surface. The non-antipodal spot, in the southern hemisphere of the star, is responsible for the hottest blackbody temperature ($0.87$~keV) for both epochs, and its semi-aperture increases from 2013 to 2016. This temperature is very close to $0.88$~keV, as found by \citet{2017MNRAS.471.1819C} when fitting \sgr~spectrum with a single hot spot.

\begin{table}[t]
%\begin{small}
\centering
\begin{tabular}{l c c}
\multicolumn{3}{c}{Best Solutions}\\
	\hline
		  & {\it D13} & {\it D16}  \\
          \hline
         $\text{GoF}$& 0.11 & 0.27 \\
		 $M~(M_\odot)$ & $1.40$ & $1.40$  \\
		 $R$~(km)& $10.97$ & $11.02 $ \\
		 $i $ & $57\degr$ & $58\degr$\\
		 $j$ & $57\degr$ & $56\degr$\\
		 $N_{\sigma}$ & $3$ & $3$ \\
		 PF & $0.31$ & $0.32$ \\
		 
		 \hline
		 $\theta_{c1}$ & $22\degr$ & $40\degr$ \\
		 $\theta_1$ & $0\degr$ &  $2\degr$ \\
		 $\phi_1$ & $0\degr$ & $351\degr$\\
		 $T_1$~(keV) & $0.6967$ & $0.2857$\\
		 \hline
		 $\theta_{c2}$ & $16\degr$ & $67\degr$ \\
		 $\theta_2$ & $180\degr$ &  $178\degr$ \\
		 $\phi_2$ & $0\degr$ & $341\degr$\\
		 $T_2$~(keV) & $0.7858$ &$0.0752$ \\
		 \hline
		 $\theta_{c3}$ & $21\degr$ & $26\degr$ \\
		 $\theta_3$ & $102\degr$ &  $117\degr$ \\
		 $\phi_3$ & $234\degr$ & $225\degr$\\
		 $T_3$~(keV) & $0.8789$ & $0.8798$\\
		 \hline
\end{tabular}
%\end{small}
\caption{List of solutions found for {\it D13} and {\it D16}. The positions of the spots can be visualized in Fig. \ref{fig:star844}.}\label{best_chromo}
\end{table}	

\subsection{Statistical considerations}
\label{statistics}

\begin{table*}[t]
%\begin{small}
\centering
%\begin{tabular}{l c c c c c c}
\begin{tabularx}{0.9\textwidth}{*{7}{>{\centering\arraybackslash}X}}
\multicolumn{7}{c}{Separated fits of D13 and D16 epochs.}\\
	\hline
		  \multicolumn{1}{l}{model}                      &  A1	&	B1	&		C1	&		A2	&		B2	&		C2    \\
        \hline
		  \multicolumn{1}{l}{data}& {\it D13} & {\it D13} & {\it D13} & {\it D16} & {\it D16} & {\it D16} \\

		  \multicolumn{1}{l}{$N_{\sigma}$}					&  2		&	2		&		3		&		2		&		2		&		3    	   \\
		  \multicolumn{1}{l}{antipodal}					&  y		&	n		&		y		&		y		&		n		&		y          \\
		  \multicolumn{1}{l}{GoF}                 &  0.29		&	0.22	&		0.11	&		0.33	&		0.35	&		0.27       \\
		  \multicolumn{1}{l}{$\sigma$}                          &  0.09		&	0.09	&		0.09	&		0.15	&		0.15	&		0.15       \\
		  \multicolumn{1}{l}{$\chi^2 (D_i)$}                    &  35.80	&	27.16	&		13.58	&		14.67	&		15.56	&		12.00      \\
		  
		  \multicolumn{1}{l}{$\chi^2_{\text{red}}$}            &  5.11		&	5.43	&		4.53	&		2.10	&		3.11	&		4.00       \\
		  \multicolumn{1}{l}{NFP}        &  9		&	11		&		13		&		9		&		11		&		13         \\
		  \multicolumn{1}{l}{NDP}              &  16		&	16		&		16		&		16		&		16		&		16         \\
		  \multicolumn{1}{l}{DoF}       &  7		&	5		&		3		&		7		&		5		&		3		   \\
\hline
\multicolumn{1}{l}{models} & A1/B1 & B1/C1 & B2/C2 \\
\hline
\multicolumn{1}{l}{F statistics} & 0.7954 &  1.4999 & 0.4444 \\
\multicolumn{1}{l}{p-value} & 	0.5012 & 0.3535 & 0.6775 \\

 %       {F statistcs}	& &	0.7954712813	&	1.499955818		&	&	&	0.4444375	\\
 %       {p-value} &	&	0.5012520865	&	0.3535612011	&	&	&	0.677557109 \\

% \hline
\end{tabularx}
%\end{small}
\caption{Acronym meanings: Number of fitting parameters (NFP), Number of Data Points (NDP), Degrees of Freedom (DoF). $\sigma$ stands for the data uncertainty. The row ``antipodal'' specifies whether models have (y) or do not have (n) two antipodal spots. The F-statistics and p-value are calculated by comparing two models as indicated by $(X1)/(X2)$. The mean value of the mass been fixed by the zeroth-run, while the radii have been kept free for both epoch fits (see Sec. \ref{Results} for details).}\label{runs_sep}
\end{table*}	
\begin{table*}[t]
%\begin{small}
\centering
%\begin{tabular}{l c c c c c c c c}
\begin{tabularx}{0.9\textwidth}{*{9}{>{\centering\arraybackslash}X}}
\multicolumn{9}{c}{Simultaneous fits of D13 and D16 with free masses and radii.}\\
\hline
\multicolumn{1}{l}{model} & 	\multicolumn{2}{c}{D} & 	\multicolumn{2}{c}{E} & 	\multicolumn{2}{c}{F}  & 	\multicolumn{2}{c}{G}  \\
\hline
\multicolumn{1}{l}{data}						&			\multicolumn{2}{c}{$D13 + D16$}				&			\multicolumn{2}{c}{$D13 + D16$}				&				\multicolumn{2}{c}{$D13 + D16$}				&				\multicolumn{2}{c}{$D13 + D16$}				\\
\multicolumn{1}{l}{$N_{\sigma}$}					& \multicolumn{2}{c}{2}					&				\multicolumn{2}{c}{2}					&					\multicolumn{2}{c}{3}					&					\multicolumn{2}{c}{3}					\\
\multicolumn{1}{l}{antipodal}					&			\multicolumn{2}{c}{y}					&				\multicolumn{2}{c}{n}					&					\multicolumn{2}{c}{y}					&					\multicolumn{2}{c}{n}					\\
\multicolumn{1}{l}{GoF}					&			\multicolumn{2}{c}{0.5882}					&				\multicolumn{2}{c}{0.5589}					&					\multicolumn{2}{c}{0.5203}					&					\multicolumn{2}{c}{0.3761}					\\
\multicolumn{1}{l}{$\chi^2$ ({\it D13+D16})}			&	\multicolumn{2}{c}{45.21}				&				\multicolumn{2}{c}{42.71}				&					\multicolumn{2}{c}{39.28}				&					\multicolumn{2}{c}{19.05}			\\

\multicolumn{1}{l}{$\chi^2_{\text{red}}$}            &				\multicolumn{2}{c}{2.83	}				&				\multicolumn{2}{c}{3.56	}				&					\multicolumn{2}{c}{4.91	}				&					\multicolumn{2}{c}{4.76	}			\\
\multicolumn{1}{l}{NFP}        &				\multicolumn{2}{c}{16	}				&				\multicolumn{2}{c}{20	}				&					\multicolumn{2}{c}{24	}				&					\multicolumn{2}{c}{28	}				\\
\multicolumn{1}{l}{NDP}              &				\multicolumn{2}{c}{32	}				&				\multicolumn{2}{c}{32	}				&					\multicolumn{2}{c}{32	}				&					\multicolumn{2}{c}{32	}				\\
\multicolumn{1}{l}{DoF}      &				\multicolumn{2}{c}{16	}				&				\multicolumn{2}{c}{12	}				&					\multicolumn{2}{c}{8	}				&					\multicolumn{2}{c}{4	}				\\

% 		F statistcs	& \multicolumn{2}{c}{}	&	\multicolumn{2}{c}{0.176}	&	\multicolumn{2}{c}{0.1746435845}	&	\multicolumn{2}{c}{1.061942257	} \\
%		p-value		&	\multicolumn{2}{c}{} &	\multicolumn{2}{c}{0.947}	&	\multicolumn{2}{c}{0.9452530984}	&	\multicolumn{2}{c}{0.4774762276	} \\
\hline
%\multicolumn{9}{c}{Simultaneous results are shown separated for {\it D13} and {\it D16}.} \\
%\hline

\multicolumn{1}{l}{$D_i$}				& {\it D13}		&		{\it D16}			&		{\it D13}		&		{\it D16}			&		{\it D13}			&		{\it D16}			&		{\it D13}			&		{\it D16}			\\
\multicolumn{1}{l}{GoF}				& 0.2413 & 0.3469	&	0.2260  &	0.3329	&	0.2044	&	0.3159	&	0.0294	&	0.3467	\\
\multicolumn{1}{l}{$\sigma$}				& 0.09	&		0.15		&		0.09	&		0.15		&		0.09		&		0.15		&		0.09		&		0.15		\\
\multicolumn{1}{l}{$\chi^2 (D_i)$}				& 29.79	&		15.42		&		27.91	&		14.80		&		25.24		&		14.04		&		3.64		&		15.41		\\
\hline
\multicolumn{1}{l}{models} & D/E & E/F & F/G \\
\hline
\multicolumn{1}{l}{F statistics} & 0.1760 & 0.1746 & 1.0619 \\
\multicolumn{1}{l}{p-value} & 0.9470 & 0.9452 & 0.4774  \\

% GoF	& 0.2413325491 & 0.3469317932	&	0.2260486582  &	0.3329843871	&	0.2044463437	&	0.3159153451	&	0.02946590883	&	0.3467393577	\\
%F statistics & p-value \\
%D/E & 0.1760 & 0.9470 \\
%E/F & 0.1746 & 0.9452 \\
%F/G & 1.0619 & 0.4774 \\
\end{tabularx}
%\end{small}
\caption{The mass and radius were free to vary and have been simultaneously
adjusted for the {D13} and {D16} epochs. The intermediate GoF and $\chi^2$ ({\it D13} and {\it D16}) for the simultaneous fits are shown in the mid-part of the table. The meaning of the acronyms and statistics are the same as in Tab. \ref{runs_sep}.}\label{runs_simult}
\end{table*}

\begin{table*}[t]
%\begin{small}
\centering
%\begin{tabular}{l c c c c}
%\begin{tabular*}{\textwidth}{@{\extracolsep{\textwidth minus\textwidth}}*5{c}@{}}
\begin{tabularx}{0.6\textwidth}{*{5}{>{\centering\arraybackslash}X}}
\multicolumn{5}{c}{Simultaneous fittings for D13 and D16. Stellar mass is fixed at $1.4 M_{\odot}$.}\\
\hline
\multicolumn{1}{l}{model} & 	\multicolumn{2}{c}{H} & 	\multicolumn{2}{c}{I} \\

\hline
\multicolumn{1}{l}{data}	&					\multicolumn{2}{c}{D13 + D16}				&			\multicolumn{2}{c}{D13 + D16}				\\
\multicolumn{1}{l}{$N_{\sigma}$}					&					\multicolumn{2}{c}{2}			&				\multicolumn{2}{c}{3}		\\
\multicolumn{1}{l}{antipodal}					&					\multicolumn{2}{c}{y}			&				\multicolumn{2}{c}{y}		\\
\multicolumn{1}{l}{GoF}					&					\multicolumn{2}{c}{0.5976}			&				\multicolumn{2}{c}{0.5272}		\\
\multicolumn{1}{l}{$\chi^2 (D13+D16)$}			&					\multicolumn{2}{c}{46.40}			&				\multicolumn{2}{c}{40.20}		\\	
\multicolumn{1}{l}{$\chi^2_{\text{red}}$}            &						\multicolumn{2}{c}{2.73}			&				\multicolumn{2}{c}{4.47}		\\	
\multicolumn{1}{l}{NFP }       &						\multicolumn{2}{c}{15}			&				\multicolumn{2}{c}{23}		\\	
\multicolumn{1}{l}{NDP}              &					\multicolumn{2}{c}{32}			&				\multicolumn{2}{c}{32}		\\	
\multicolumn{1}{l}{DoF}       &							\multicolumn{2}{c}{17}			&				\multicolumn{2}{c}{9}		\\
% F statistcs & \multicolumn{2}{c}{}  & \multicolumn{2}{c}{0.173507462686567} \\
% p-value & \multicolumn{2}{c}{}  & \multicolumn{2}{c}{0.989182249918315} \\
\hline
% \multicolumn{5}{c}{Simultaneous results are shown separated for {\it D13} and {\it D16}.} \\
%\hline					
\multicolumn{1}{l}{$D_i$}				&		D13			&		D16			&			D13  & D16				\\
\multicolumn{1}{l}{GoF}				&		0.2511	&	0.3465	&	0.2122		&		0.3150			\\
\multicolumn{1}{l}{$\sigma$}				&		0.09			&	0.15			&	0.09		&		0.15			\\
\multicolumn{1}{l}{$\chi^2 (D_i)$}				& 		31.00			&	15.40			&	26.20		&		14.00			\\
\hline
\multicolumn{1}{l}{models} & H/I \\
\hline
\multicolumn{1}{l}{F statistics} & 0.1735  \\
\multicolumn{1}{l}{p-value} & 0.9891 \\
%\multicolumn{1}{l}{H/I} & 0.1735 & 0.9891 & & \\
%\end{tabular*}
\end{tabularx}
%\end{small}
\caption{The meaning of the acronyms are the same as in Tab. \ref{runs_sep}. The middle part refers to the intermediate GoF and $\chi^2$ for the simultaneous fits, as in Tab. \ref{runs_simult}.}\label{runs_simult2}
\end{table*}

Given that some macroscopic aspects of the star should not change significantly from one period to the other, important conclusions could already be reached from one dataset alone, for example {\it D13}. Clearly, three hot spots can fit better the data than two hot spots. This can be seen by their goodness-of-fit per degree of freedom ($\text{GoF}/\text{DoF}$), as present in Figs.  \ref{two-hot-spots} and \ref{fig:gen837} and in Tab. \ref{runs_sep}. However, for meaningful fit comparisons we calculate the standard reduced $\chi^2$, $\chi^2_{\mbox{red}}\coloneqq\text{GoF}/ (\sigma^2 \text{DoF})=\chi^2/\text{DoF}$ ($\sigma$ is the normalized error bar of the measurements).
As clear from Tab. \ref{runs_sep}, one can see that $\chi^2_{\mbox{red}}=2-6$ for both datasets. 
A possible interpretation of the large values of $\chi^2_{\mbox{red}}$ is an over-fitting due to the small number of data points (resulting in a small DoF).
We have also performed the F-test between nested models. The p-values of these statistics suggest that there is not a preferred model.
This is not surprising given the large number of parameters when compared to the data (small number of degrees of freedom).

In order to increase the number of degrees of freedom, we have also attempted to fit the data in other ways. We have assumed the case where the {\it D13} and {\it D16} datasets are fit simultaneously for certain parameters. Our results are summarized in Tabs.~\ref{runs_simult} (for free fitting masses and radii) and \ref{runs_simult2} (free fitting radii and fixed mass at $1.4$ M$_{\odot}$). As one can clearly see, none case led to a preferred hot spot scenario. For instance, the $\chi^2_{\mbox{red}}$ found are as large as before, which is yet a consequence of the very small number of observational data for \sgr. One might wonder what is the minimum amount of data points needed to reach more stringent results. As the goodnesses of fit of Tab.~\ref{runs_simult} already suggest, assume that this hypothesized case still leads to $GoF\approx 0.6$ to the simultaneous fit. Then, it follows that $\chi^2_{\mbox{red}}\approx 1.05$ would be reached when the degrees of freedom are approximately 70. This is much larger than our \ssgr~data. We come back to this issue in Sec. \ref{discussion}. 

\section{Additional systematic uncertainties to $M$ and $R$}
\label{uncertainties}

Care should be taken when extracting physical information from pure blackbody emission models. The processes responsible for radiation emission in SGRs/AXPs are still largely unknown. They may be related to the presence of an atmosphere, although with properties quite different from those of standard atmospheres around passively cooling NSs, or even arise from a condensed surface. In both cases, the spectrum is expected to be thermal but not necessarily blackbody-like~\citep[see, e.g.,][and references therein]{2014PhyU...57..735P}. In the case of NSs, one can expect that the emitting layers are comprised of just one, lightest available, chemical element because heavier elements sink into deeper layers due to the immense NS gravitational field.

Several works have addressed the problem of modelling the radiation transport in magnetized NS atmospheres. \citet{1992A&A...266..313S} were the first to perform detailed calculations of radiation spectra emerging from
strongly magnetized NS photospheres, 
for the case of a fully ionized plasma.
Besides, they have created a database of magnetic hydrogen spectra~\citep[see also][and references therein]{2001MNRAS.327.1081H,2007MNRAS.375..821H} and have shown that the spectra of magnetic hydrogen and helium atmospheres are softer than the nonmagnetic ones, but harder than the blackbody spectrum with the same temperature. 
Thus, if an amount of hydrogen is present in the outer layers (e.g., because of accretion of the interstellar matter), one can expect a pure hydrogen atmosphere. The latter can lead to much harder spectra 
in the \text{Wien tail} than the blackbody spectrum, because hotter deep layers are seen at high frequencies, where the spectral opacity is lower~\citep{1996ApJ...472L..33P}. In this case, the best-fit effective temperature of the atmosphere is considerably lower than the blackbody temperature, whereas the $R/D$ ratio is larger than the one for the blackbody fit. Therefore, models that go beyond blackbody assumptions could have an important influence on \sgr mass and radius constraints.

A crude way of estimating further uncertainties to our $M$ and $R$ results due to the presence of atmospheres (e.g., hydrogen) could be as follows. One could average out the different hot spot temperatures in the {\it D13} and {\it D16} datasets and find a representative temperature and an uncertainty to them. With this uncertainty, one could estimate a range of wavelengths around the one for the maximum flux, $\lambda_{max}$ (the most relevant wavelength for a given temperature), and then use known atmospheric models \citep{2007PhRvL..98g1101P} to find the largest change of the flux (with respect to the blackbody) for this wavelength interval. Finally, by extrapolating these results, one gets the flux change estimates to our case.  
Using the spots' temperatures from
Tab.~\ref{best_chromo}, one has that a representative value for them is $0.6\pm 0.3$ keV ($7.0\pm 0.3\times 10^6$~K). \footnote{If one assumes that the flux of the hot spots is around 10 times larger than the one from the star's surface \citep{2001ApJ...559..346D}, then the mean hot spot temperature should be around twice as large as the star's surface. This allows us to conclude that our fit parameters are in good agreement with independent fits of surface temperatures and magnetic fields of stars \citep{2007PhRvL..98g1101P} since the surface dipolar magnetic field of \sgr would be around $2\times 10^{14}$ G \citep{2015MNRAS.449.2685C}.} For the above hot spot temperature uncertainty, one then expects the relevant wavelengths to range from $(2/3)\lambda_{max}$ to $2\lambda_{max}$. From Fig.~$6$ of \citet{2007MNRAS.375..821H} (or Fig.~$1$ of \citealp{2009A&A...500..891S}), it thus follows that hydrogen atmospheres of isolated magnetized stars should lead to a maximum difference in flux of approximately $20\%$ when compared to blackbody results.  
If now one goes back to the expression of the flux and takes it as a function of $M$ and $R$, it follows that a $20\%$ change of it leads to a maximum uncertainty of approximately $7\%$ to the radius and a $5\%$ uncertainty to the mass with respect to blackbody outcomes. In order to reach these differences, we have taken $I_p=I_s\approx 0.015$ as a representative value. We note that the above changes could either increase or decrease the mean mass and radius of \sgr.

Another source of uncertainty to our blackbody-based results is light beaming. This is specially the case for systems with high magnetic fields \citep{2009A&A...500..891S,2001ApJ...559..346D}, as is very likely the case of \ssgr\,\citep{2015MNRAS.449.2685C}. Besides the plasma present in the magnetosphere, the presence of an accretion column itself could lead the emission from hot spots to be beamed \citep{2001ApJ...559..346D}. When compared to isotropic emission models, beaming could change pulsed fractions substantially \citep{2001ApJ...559..346D}. One could crudely estimate additional uncertainties to our model in the following way. The averaged semi-aperture angle from our hot spots is $\bar{\theta}_{c}\approx 32\degr$ (see Table~\ref{best_chromo}). From our model, the \ssgr\, pulsed fraction is approximately $0.3$. Assuming that the hot spots could have a flux around 10 times larger than the star's surface \citep{2001ApJ...559..346D}, from Fig. $4$ of \citet{2001ApJ...559..346D}, one sees that the most appropriate beaming index in this case should be $n=1$ ($I\propto \cos^n\alpha$) and changes in the maximum to minimum flux ratio could be $65\%$ (pulsed fraction going from $0.1$, the maximum in the isotropic case \citep{2001ApJ...559..346D}, to $0.3$, the inferred one from our analysis of \ssgr). This means, crudely speaking, that the flux could change around $30\%$ from a pure blackbody. In terms of differences to macroscopic parameters, following the procedure laid out before for atmospheres, we find that beaming leads to a maximum difference of $6\%$ to the mass and $10\%$ to the radius. We stress that this is very model and parameter dependent and it is not excluded larger or smaller corrections to blackbody outcomes. We comment further on beaming in the discussion section.

All the above systematic uncertainties indicate that, so far, it is not possible to make predictions for the mass and radius of \sgr as precise as one would wish. Combining the above models, systematic modelling uncertainties could lead the radius and the mass to change by up to $20\%$ and $10\%$, respectively. However, a clear aspect from our simple analysis is that fits with three hot spots resulted in smaller $\text{GoF}$, meaning that they are more statistically relevant than two hot spots. We discuss possible interpretations of that in the next section.

\section{Discussions and conclusions}
\label{discussion}

Chandra X-ray data have been used to constrain SGR J1745--2900 properties by means of  genetic algorithm techniques. 
From \ssgr\, X-ray light-curve and pulsed fraction and the assumption that they come from stellar hot spots of any size, temperature and stellar position, 
fits have been made attempting to reproduce as best as possible the data. We took into account relativistic effects such as light bending and we have ignored the effects of stellar rotation, well supported by the \ssgr\, long rotation period ($3.76$~s). In this first approach, we have also ignored atmospheric effects and beaming on the fits.
Global and split into two epochs data have been investigated for uncertainty estimations and precise parameter extractions. 

Although fits with three hot spots lead to better-than-two GoFs, statistical considerations have shown that both models are equivalent. This is due to the limitation of the observational data itself, which severely decreases the degrees of freedom of the system for the models. Even though the resultant statistics is poor in any case, one could interpret the above-mentioned ambiguity as a suggestion that a multipolar structure in \sgr should not be excluded. This comes from the fact that at least one model we have analyzed is a reasonable first-order description to neutron stars. Indeed, this should be the case for dipolar models since braking index measurements for pulsars are not too far from three~\citep[see e.g.,][]{2016ApJ...823...97C,2016JCAP...07..023D,2016ApJ...831...35D,2016EPJC...76..481D,2017EPJC...77..350D} and some properties of SGRs/AXPs would need strong dipolar fields~\citep{2017A&A...599A..87C}. Thus, if two hot spots are reasonable at the surfaces of stars and they are statistically equivalent to three hot spots, one should not disregard the latter (or other situations with more hot spots) in modelling NS light-curves. This has indeed been shown to be the case of pulsar \psr, which strengthens even further the suggestions of our statistical analysis for \sgr. We leave for future work investigations of light-curves of neutron stars with more data points using the GA techniques developed here. In particular, we plan to investigate \psr, given that the hot spot configuration found for it is very different from what expected in the dipolar case \citep[see][]{Bilous_2019,Riley_2019,Raaijmakers_2019,Miller_2019,Bogdanov_2019,Bogdanov_2019_,Guillot_2019}.

Regarding the normalized flux fits, some words are in order. Firstly, we have not fitted both datasets entirely independently. We have taken the mean mass from our zeroth run (with all datasets run simultaneously; see Fig.~\ref{fig:HistMass861}) in order to minimize the computation time of other parameters. This is reasonable because \sgr is an isolated NS. We have not taken the mean value of the radius from our zeroth run, but we have treated it as a free parameter in the {\it D13} and {\it D16} fits. However, we expect them, as well as the inclination angles {\it i} and {\it j}, to remain almost the same, as indeed happened to many populations, and that has been used as our criterion for selecting ``the best'' solution (see Tab. \ref{best_chromo}). \footnote{The GA we have made use of has a mutation parameter to prevent solutions from getting stuck in a false minimum. We have taken it to be $0.1$, meaning that in every generation $10\%$ of the population suffers mutation. Besides that, we have used many initial populations and have stopped running generations when the best solution (minimum of $\chi^2$) had been the same for many successive generations (around 1000). Not all populations converged to the same solution, but we selected the physical one as the best of those with the same macroscopic parameters for both datasets.} This shows consistency in our simple model. 
Nonetheless, the fit of the last points of the {\it D16} dataset indicates that the model is not entirely appropriate.
This could be due to several reasons, one of them being the small amount of data itself \citep[see][]{2017MNRAS.471.1819C}. Another reason would be that we have modelled the data in a very simple way, forcing both epochs to be described equally, and important effects might have been left out. An example of that could be significant changes of \sgr atmospheric conditions from one epoch to the other. A sharp change of the beaming might also take place, meaning that accretion columns could change their properties due to an outburst. Indeed, it could rearrange or disturb the atmosphere of the magnetized NS, and as a result the flux could change non-negligibly. Thus, better fits could raise if different atmospheric models are taken for the epochs analyzed, which we have not done in this first analysis. We plan to elaborate on the above in future works.

The uncertainties to $M$ and $R$, coming from our zeroth run, should be taken just as indicative. Systematic uncertainties due to different models could also be relevant. We have investigated some of them and it seems that
atmospheric models and beaming could play an important role into more realistic uncertainties to the parameters. Rough estimates suggest that variations of the flux with respect to our model are around $50\%$, meaning an additional $20\%$ ($10\%$) radius (mass) uncertainty to \sgr's mean blackbody outcomes. However, it is important to bear in mind that models for NS atmospheres are still debatable and blackbody results could give us interesting insights for testing them more precisely.

\begin{figure*}[h]
\centering
\includegraphics[width=1\hsize,clip]{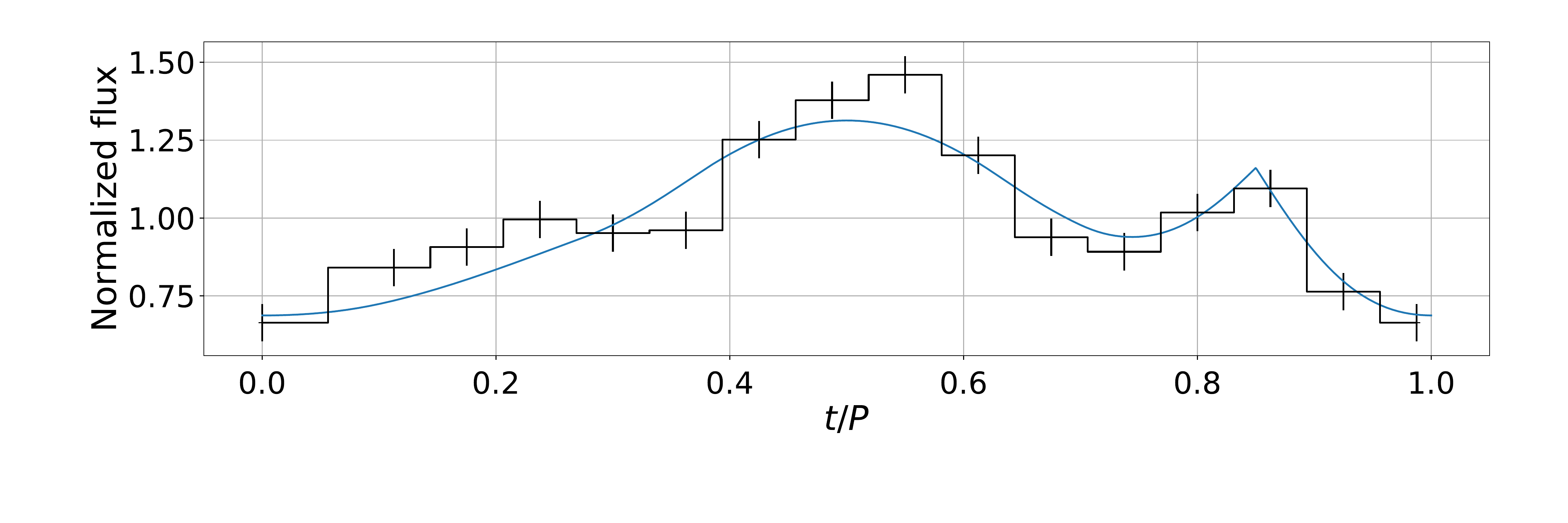}
\includegraphics[width=1\hsize,clip]{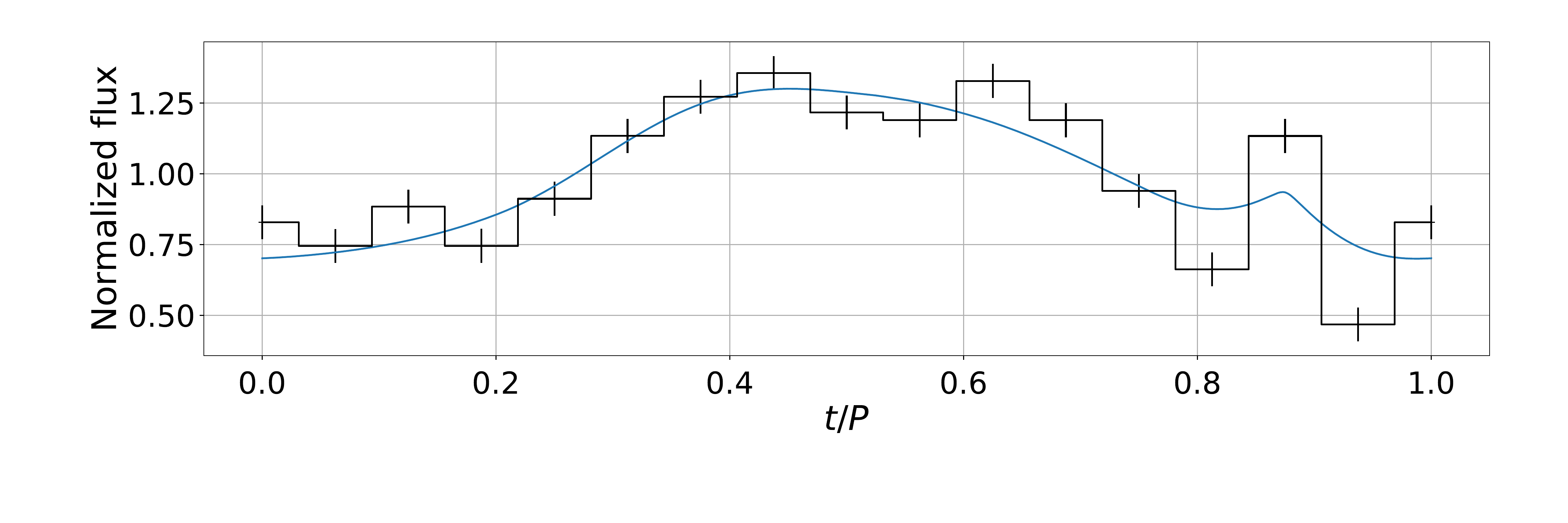}
\caption{Upper panel: D13's fitting for three spots. The mass is $1.4~M_{\odot}$ and two spots are antipodal, while the third one is free. The parameters found were $R=10.97~$km, $j=57\degr$, $i=57\degr$,$T_{1}= 0.6967~$keV, $T_{2}=0.7858~$keV, $T_{3}=0.8789~$keV. GoF = $0.11$ and the number of degrees of freedom here is 3 (the mass has been fixed by our zeroth run), which implies that $\text{GoF}/\text{DoF} = 0.037$. Bottom panel: D16's fitting for three spots. Same mass and spot configurations as the D13 set. The parameters found were $R=11.02~$km , $j=58\degr$, $i=56\degr$,$T_{1}= 0.2857~$keV, $T_{2}=0.0752~$keV, $T_{3}=0.8798~$keV. GoF = $0.27$ ($\text{GoF}/\text{DoF} = 0.09$).}
\label{fig:gen837}
\end{figure*}

\begin{figure*}[t]
\centering
\includegraphics[width=0.5\hsize,clip]{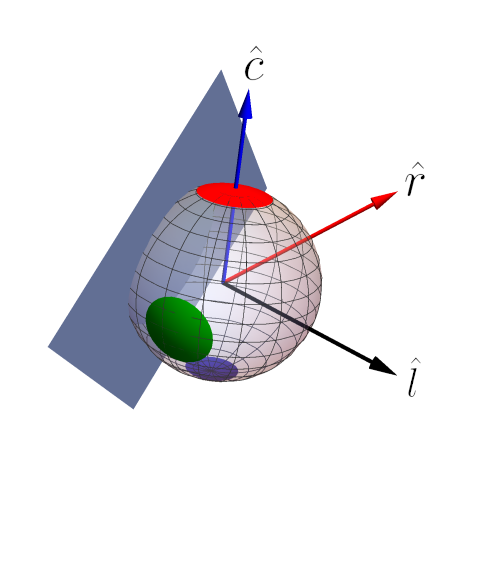}\includegraphics[width=0.5\hsize,clip]{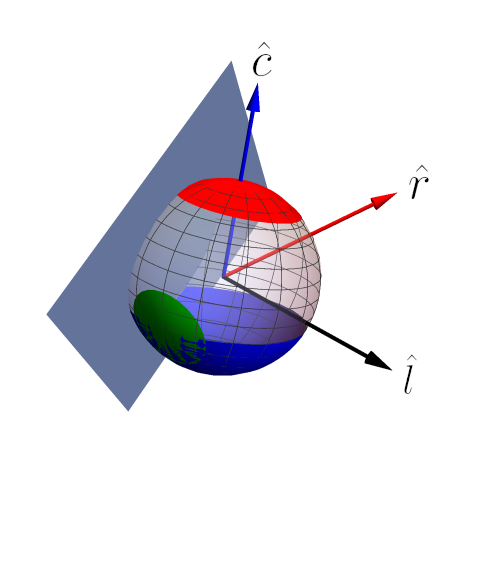}\caption{Left panel: D13's three spots positions.  $T_1=0.6967~$keV (north pole spot), $T_2=0.7858~$keV (south pole spot), $T_3=0.8789~$keV (non-antipodal--or southern hemisphere--spot).
Right panel: D16's three spots positions. ,$T_1=0.2857~$keV (north pole spot), $T_2=0.0752~$keV (south pole spot), $T_3=0.8798~$keV (non-antipodal spot). The arrows shown are the $\hat{\mathbf l}$ (LOS), around the star's equatorial plane, $\hat{\mathbf{c}}$ (polar cap axis), crossing the north pole, and $\hat{\mathbf r}$ (rotation axis), the remaining arrow in the northern hemisphere. A plane is drawn as a reference to the maximum angle $\theta_{F}$ from which the observer cannot receive signals anymore.}
\label{fig:star844}
\end{figure*}

We now make a few comments regarding the case the surface of \sgr has three hot spots. The hot spots in Fig.~\ref{fig:star844}, in the light of the \citet{Gourgouliatos_2017} results, could be interpreted as follows. First of all, the magnetic field at the stellar surface in both data sets seems to be far from axially symmetric because of the better fits coming from three hot spots. For the {\it D13} set, the presence of the non-antipodal spot (southern hemisphere), whose size is comparable to the antipodal spots (north and south poles), suggests that the toroidal field should be relevant. Indeed, purely dipolar models would lead to spot areas of the order of the polar cap area $A_{\rm pc} = \pi R_{\rm pc}^2$, where $R_{\rm pc} = \sqrt{2\pi R^3/(c P)}$ \citep[see, e.g.,][]{1975ApJ...196...51R,1977ApJ...214..598C,1993ApJ...402..264C}, and, for a NS with $R=11$~km and $P=3.76$~s, $A_{\rm pc}\approx 0.023$~km$^2$, much smaller than the areas of the spots in Fig.~\ref{fig:star844}. This clearly indicates that the magnetic field of SGR J1754–-2900 is very different from a dipolar configuration. According to \citet{Gourgouliatos_2017}, a very localized spot ($\approx 1$~km) implies a very specific configuration where $99\%$ of the energy is in the toroidal field. However, smaller toroidal energy budgets lead to more extended magnetic zones at the stellar surface and, as a consequence, an extended hot region \citep{Gourgouliatos_2017}. Therefore, our results suggest that SGR J1754–-2900 has a complex multipolar magnetic field structure, with a relevant toroidal component for both {\it D13} and {\it D16} data sets (not overwhelmingly dominant, though, because the hot spots are not small). Indeed, the variability of the spindown rate of SGR J1745--2900 implies that its characteristic age ($\approx 4.3$~kyr) is accurate to its real age up to one order of magnitude, meaning it would be a young source and hence it might have a quite complex magnetic field structure. In addition, the association of some SGRs/AXPs with supernova remnants suggests that the ages of these sources are typically $\leq 10^4$--$10^5$~yr~\citep[see, e.g.,][]{2017ARA&A..55..261K}.

The variation of the spots temperatures and sizes from one epoch to the other is pronounced. One might interpret these results as due to thermal conduction and temperature gradients on the stellar surface. This seems reasonable given the very large electric conductivity of the star, which would also imply a very large thermal conductivity, and so very small timescales for temperature variations. The temperature change of the spot at the north pole might be associated with its expansion, triggered by temperature gradients, and standard cooling processes. The significant temperature decrease of the south pole hot spot might also be due to its large increase, possibly triggered by similar reasons as to what happened to the north pole hot spot. However, the temperature change in the non-antipodal spot has been practically zero, and that might be related to its partial overlap with the south pole hot spot.

Apart from temperature values of some of the hot spots of \ssgr, our results contrast with those of \citet{2017MNRAS.471.1819C} for the same source and data. Firstly, we have taken two and three hot spots, while they assume just a single one. Secondly, we have found that the sizes of the spots increase from 2013 to 2016, while the opposite happens to their single spot. In their case, the spot shrinking was important to explain the increase of the pulsed fraction. In our case, the increase of the pulsed fraction might be explained with the large temperature changes of some spots from one epoch to the other. Due to the relevance of hot spot size evolution to physical processes taking place in stars \citep{2017MNRAS.471.1819C}, we leave precise analyses thereof in light of our results to the carried out elsewhere.

We stress an important point of our analysis. One can see from the bottom panel of Fig.~\ref{fig:gen837} that our best fit to the normalized flux has not been so good for the last 2016 data points. This means that our pulsed fraction increase is not as pronounced (see Table \ref{best_chromo}) as the observed one [from approximately 0.35 to 0.58 \citep{2017MNRAS.471.1819C}]. We have tried to enhance the 2016 fit with three free hot spots on the stellar surface, but no better results have been found. Since in this case the number of free parameters is the same as the data points for each set, we have kept analyses with three hot spots where two of them are antipodal, which naturally have less parameters than data. Thus, it is still pending ways to enhance the fit of the last data points of the 2016 light-curve of \ssgr.

We have performed a light-curve and pulsed fraction X-ray data analysis of \ssgr\, without assuming any specific nuclear EOS. The data analysis based on the blackbody model alone indicates that SGR J1745--2900 has as the most likely mass the canonical NS mass $M= 1.4~M_\odot$, and it should have a corresponding radius  
$R_{1.4}\approx 9.4$--$12.3$~km. This result obtained from electromagnetic data agrees with recent constraints obtained from gravitational wave observations that lead to $R_{1.4}\lesssim 13.5$~km for hadronic stars  \citep{2018PhRvL.120q2703A,2018PhRvL.121i1102D,2018PhRvL.121p1101A,2018PhRvL.120z1103M}. The above values would disfavour relativistic mean-field theory models, which usually lead to $R_{1.4}$ larger than $13.5$~km \citep{2016PhRvC..94c5804F}. Some Skyrme models (see, for instance, Fig. $7$ of \citealp{2016PhRvC..94c5804F}, were models should have $R_{1.4}$ in the range of $11.5-13.5$~km), as well as the MPA1, APR and WFF parameterizations (see their $R_{1.4}$ in \citealp{2009PhRvD..79l4032R}), among other EOS, especially stiffer, seem to be favoured by our analysis. However, the systematic modelling uncertainties that we have pointed out before significantly weaken the above EOS constraints, and no definite conclusion can be reached so far; this might be mitigated just when precise emissions models are analyzed or when more data is collected. Finally, it remains open the question whether SGR J1745--2900 could be a hybrid star since many of the hybrid EOS would lead to a third-family of NSs which would satisfy our light-curve constraints (see, for instance, \citealp{2019A&A...622A.174S,2018PhRvD..97h4038P} and references therein).

Summing up, we have carried out fits of the light-curve of \sgr using genetic algorithms techniques. Although the observational data of \sgr 
is not enough to achieve stringent statistical conclusions, our analysis gave us important hints on magnetic fields of SGRs/AXPs. The fact that two or three hot spots could equally describe the data of \sgr suggests that in NS cases with more observations one should not disregard a multipolar structure of their magnetic fields.

\section*{Acknowledgements}
We thank the anonymous referee for important suggestions which have improved our work. We would like to thank F. C. Zelati for providing us with the light-curve data. We are grateful to Professor John Boguta for insightful discussions on the nuclear EOS and relativistic mean-field theory models. R.C.R.L. acknowledges the support of Funda\c c\~ao de Amparo \`a Pesquisa e Inova\c c\~ ao do Estado de Santa Catarina (FAPESC) under grant No. 2017TR1761. J.G.C. is likewise grateful to the support of Conselho Nacional de Desenvolvimento Cient\'ifico e Tecnol\'ogico - CNPq (421265/2018-3 and 305369/2018-0).
J.P.P. acknowledges the financial support given by Funda\c c\~ao de Amparo \`a Pesquisa do Estado de S\~ao Paulo (FAPESP) (2015/04174-9 and 2017/21384-2), and the Polish National Science Centre under grant No. 2016/22/E/ST9/00037.
C.V.R. is grateful to grants from CNPq (303444/2018-5) and FAPESP (2013/26258-4).

\newpage
\bibliographystyle{apj}
\bibliography{ref}

\begin{thebibliography}{52}
\expandafter\ifx\csname natexlab\endcsname\relax\def\natexlab#1{#1}\fi

\bibitem[{{Abbott} {et~al.}(2018){Abbott}, {Abbott}, {Abbott}, {Acernese},
  {Ackley}, {Adams}, {Adams}, {Addesso}, {Adhikari}, {Adya}, \&
  et~al.}]{2018PhRvL.121p1101A}
{Abbott}, B.~P., {Abbott}, R., {Abbott}, T.~D., {et~al.} 2018, Physical Review
  Letters, 121, 161101

\bibitem[{{Annala} {et~al.}(2018){Annala}, {Gorda}, {Kurkela}, \&
  {Vuorinen}}]{2018PhRvL.120q2703A}
{Annala}, E., {Gorda}, T., {Kurkela}, A., \& {Vuorinen}, A. 2018, Physical
  Review Letters, 120, 172703

\bibitem[{{Beloborodov}(2002)}]{2002ApJ...566L..85B}
{Beloborodov}, A.~M. 2002, \apjl, 566, L85

\bibitem[{{Belvedere} {et~al.}(2015){Belvedere}, {Rueda}, \&
  {Ruffini}}]{2015ApJ...799...23B}
{Belvedere}, R., {Rueda}, J.~A., \& {Ruffini}, R. 2015, \apj, 799, 23

\bibitem[{Bilous {et~al.}(2019)Bilous, Watts, Harding, Riley, Arzoumanian,
  Bogdanov, Gendreau, Ray, Guillot, Ho, \& Chakrabarty}]{Bilous_2019}
Bilous, A.~V., Watts, A.~L., Harding, A.~K., {et~al.} 2019, The Astrophysical
  Journal, 887, L23

\bibitem[{Bogdanov {et~al.}(2019{\natexlab{a}})Bogdanov, Guillot, Ray, Wolff,
  Chakrabarty, Ho, Kerr, Lamb, Lommen, Ludlam, Milburn, Montano, Miller,
  Bauböck, Özel, Psaltis, Remillard, Riley, Steiner, Strohmayer, Watts, Wood,
  Zeldes, Enoto, Okajima, Kellogg, Baker, Markwardt, Arzoumanian, \&
  Gendreau}]{Bogdanov_2019}
Bogdanov, S., Guillot, S., Ray, P.~S., {et~al.} 2019{\natexlab{a}}, The
  Astrophysical Journal, 887, L25

\bibitem[{Bogdanov {et~al.}(2019{\natexlab{b}})Bogdanov, Lamb, Mahmoodifar,
  Miller, Morsink, Riley, Strohmayer, Tung, Watts, Dittmann, Chakrabarty,
  Guillot, Arzoumanian, \& Gendreau}]{Bogdanov_2019_}
Bogdanov, S., Lamb, F.~K., Mahmoodifar, S., {et~al.} 2019{\natexlab{b}}, The
  Astrophysical Journal, 887, L26

\bibitem[{{Chen} \& {Ruderman}(1993)}]{1993ApJ...402..264C}
{Chen}, K., \& {Ruderman}, M. 1993, \apj, 402, 264

\bibitem[{{Cheng} \& {Ruderman}(1977)}]{1977ApJ...214..598C}
{Cheng}, A.~F., \& {Ruderman}, M.~A. 1977, \apj, 214, 598

\bibitem[{{Cipolletta} {et~al.}(2015){Cipolletta}, {Cherubini}, {Filippi},
  {Rueda}, \& {Ruffini}}]{2015PhRvD..92b3007C}
{Cipolletta}, F., {Cherubini}, C., {Filippi}, S., {Rueda}, J.~A., \& {Ruffini},
  R. 2015, \prd, 92, 023007

\bibitem[{{Coelho} {et~al.}(2017){Coelho}, {C{\'a}ceres}, {de Lima},
  {Malheiro}, {Rueda}, \& {Ruffini}}]{2017A&A...599A..87C}
{Coelho}, J.~G., {C{\'a}ceres}, D.~L., {de Lima}, R.~C.~R., {et~al.} 2017,
  \aap, 599, A87

\bibitem[{{Coelho} {et~al.}(2016){Coelho}, {Pereira}, \& {de
  Araujo}}]{2016ApJ...823...97C}
{Coelho}, J.~G., {Pereira}, J.~P., \& {de Araujo}, J. C.~N. 2016, \apj, 823, 97

\bibitem[{{Coti Zelati} {et~al.}(2015){Coti Zelati}, {Rea}, {Papitto},
  {Vigan{\`o}}, {Pons}, {Turolla}, {Esposito}, {Haggard}, {Baganoff}, {Ponti},
  {Israel}, {Campana}, {Torres}, {Tiengo}, {Mereghetti}, {Perna}, {Zane},
  {Mignani}, {Possenti}, \& {Stella}}]{2015MNRAS.449.2685C}
{Coti Zelati}, F., {Rea}, N., {Papitto}, A., {et~al.} 2015, \mnras, 449, 2685

\bibitem[{{Coti Zelati} {et~al.}(2017){Coti Zelati}, {Rea}, {Turolla}, {Pons},
  {Papitto}, {Esposito}, {Israel}, {Campana}, {Zane}, {Tiengo}, {Mignani},
  {Mereghetti}, {Baganoff}, {Haggard}, {Ponti}, {Torres}, {Borghese}, \&
  {Elfritz}}]{2017MNRAS.471.1819C}
{Coti Zelati}, F., {Rea}, N., {Turolla}, R., {et~al.} 2017, \mnras, 471, 1819

\bibitem[{{De} {et~al.}(2018){De}, {Finstad}, {Lattimer}, {Brown}, {Berger}, \&
  {Biwer}}]{2018PhRvL.121i1102D}
{De}, S., {Finstad}, D., {Lattimer}, J.~M., {et~al.} 2018, Physical Review
  Letters, 121, 091102

\bibitem[{{de Araujo} {et~al.}(2016{\natexlab{a}}){de Araujo}, {Coelho}, \&
  {Costa}}]{2016JCAP...07..023D}
{de Araujo}, J. C.~N., {Coelho}, J.~G., \& {Costa}, C.~A. 2016{\natexlab{a}},
  JCAP, 2016, 023

\bibitem[{{de Araujo} {et~al.}(2016{\natexlab{b}}){de Araujo}, {Coelho}, \&
  {Costa}}]{2016ApJ...831...35D}
---. 2016{\natexlab{b}}, \apj, 831, 35

\bibitem[{{de Araujo} {et~al.}(2016{\natexlab{c}}){de Araujo}, {Coelho}, \&
  {Costa}}]{2016EPJC...76..481D}
---. 2016{\natexlab{c}}, European Physical Journal C, 76, 481

\bibitem[{{de Araujo} {et~al.}(2017){de Araujo}, {Coelho}, \&
  {Costa}}]{2017EPJC...77..350D}
---. 2017, European Physical Journal C, 77, 350

\bibitem[{{DeDeo} {et~al.}(2001){DeDeo}, {Psaltis}, \&
  {Narayan}}]{2001ApJ...559..346D}
{DeDeo}, S., {Psaltis}, D., \& {Narayan}, R. 2001, \apj, 559, 346

\bibitem[{{Fortin} {et~al.}(2016){Fortin}, {Provid{\^e}ncia}, {Raduta},
  {Gulminelli}, {Zdunik}, {Haensel}, \& {Bejger}}]{2016PhRvC..94c5804F}
{Fortin}, M., {Provid{\^e}ncia}, C., {Raduta}, A.~R., {et~al.} 2016, \prc, 94,
  035804

\bibitem[{{Gendreau} {et~al.}(2016){Gendreau}, {Arzoumanian}, {Adkins},
  {Albert}, {Anders}, {Aylward}, {Baker}, {Balsamo}, {Bamford}, {Benegalrao},
  {Berry}, {Bhalwani}, {Black}, {Blaurock}, {Bronke}, {Brown}, {Budinoff},
  {Cantwell}, {Cazeau}, {Chen}, {Clement}, {Colangelo}, {Coleman},
  {Coopersmith}, {Dehaven}, {Doty}, {Egan}, {Enoto}, {Fan}, {Ferro}, {Foster},
  {Galassi}, {Gallo}, {Green}, {Grosh}, {Ha}, {Hasouneh}, {Heefner}, {Hestnes},
  {Hoge}, {Jacobs}, {J{\o}rgensen}, {Kaiser}, {Kellogg}, {Kenyon}, {Koenecke},
  {Kozon}, {LaMarr}, {Lambertson}, {Larson}, {Lentine}, {Lewis}, {Lilly},
  {Liu}, {Malonis}, {Manthripragada}, {Markwardt}, {Matonak}, {Mcginnis},
  {Miller}, {Mitchell}, {Mitchell}, {Mohammed}, {Monroe}, {Montt de Garcia},
  {Mul{\'e}}, {Nagao}, {Ngo}, {Norris}, {Norwood}, {Novotka}, {Okajima},
  {Olsen}, {Onyeachu}, {Orosco}, {Peterson}, {Pevear}, {Pham}, {Pollard},
  {Pope}, {Powers}, {Powers}, {Price}, {Prigozhin}, {Ramirez}, {Reid},
  {Remillard}, {Rogstad}, {Rosecrans}, {Rowe}, {Sager}, {Sanders}, {Savadkin},
  {Saylor}, {Schaeffer}, {Schweiss}, {Semper}, {Serlemitsos}, {Shackelford},
  {Soong}, {Struebel}, {Vezie}, {Villasenor}, {Winternitz}, {Wofford},
  {Wright}, {Yang}, \& {Yu}}]{2016SPIE.9905E..1HG}
{Gendreau}, K.~C., {Arzoumanian}, Z., {Adkins}, P.~W., {et~al.} 2016, Society
  of Photo-Optical Instrumentation Engineers (SPIE) Conference Series, Vol.
  9905, {The Neutron star Interior Composition Explorer (NICER): design and
  development}, 99051H

\bibitem[{Gourgouliatos \& Hollerbach(2017)}]{Gourgouliatos_2017}
Gourgouliatos, K.~N., \& Hollerbach, R. 2017, The Astrophysical Journal, 852,
  21

\bibitem[{Guillot {et~al.}(2019)Guillot, Kerr, Ray, Bogdanov, Ransom, Deneva,
  Arzoumanian, Bult, Chakrabarty, Gendreau, Ho, Jaisawal, Malacaria, Miller,
  Strohmayer, Wolff, Wood, Webb, Guillemot, Cognard, \&
  Theureau}]{Guillot_2019}
Guillot, S., Kerr, M., Ray, P.~S., {et~al.} 2019, The Astrophysical Journal,
  887, L27

\bibitem[{{Ho} {et~al.}(2007){Ho}, {Kaplan}, {Chang}, {van Adelsberg}, \&
  {Potekhin}}]{2007MNRAS.375..821H}
{Ho}, W.~C.~G., {Kaplan}, D.~L., {Chang}, P., {van Adelsberg}, M., \&
  {Potekhin}, A.~Y. 2007, \mnras, 375, 821

\bibitem[{{Ho} \& {Lai}(2001)}]{2001MNRAS.327.1081H}
{Ho}, W. C.~G., \& {Lai}, D. 2001, \mnras, 327, 1081

\bibitem[{{Hu} {et~al.}(2019){Hu}, {Ng}, \& {Ho}}]{2019MNRAS.485.4274H}
{Hu}, C.-P., {Ng}, C.~Y., \& {Ho}, W. C.~G. 2019, \mnras, 485, 4274

\bibitem[{{Kaspi} \& {Beloborodov}(2017)}]{2017ARA&A..55..261K}
{Kaspi}, V.~M., \& {Beloborodov}, A.~M. 2017, Annual Review of Astronomy and
  Astrophysics, 55, 261

\bibitem[{Kennea {et~al.}(2013)Kennea, Burrows, Kouveliotou, Palmer,
  Gö{\u{g}}ü{\c{s}}, Kaneko, Evans, Degenaar, Reynolds, Miller, Wijnands,
  Mori, \& Gehrels}]{Kennea_2013}
Kennea, J.~A., Burrows, D.~N., Kouveliotou, C., {et~al.} 2013, The
  Astrophysical Journal, 770, L24

\bibitem[{Miller {et~al.}(2019)Miller, Lamb, Dittmann, Bogdanov, Arzoumanian,
  Gendreau, Guillot, Harding, Ho, Lattimer, Ludlam, Mahmoodifar, Morsink, Ray,
  Strohmayer, Wood, Enoto, Foster, Okajima, Prigozhin, \& Soong}]{Miller_2019}
Miller, M.~C., Lamb, F.~K., Dittmann, A.~J., {et~al.} 2019, The Astrophysical
  Journal, 887, L24

\bibitem[{{Mori} {et~al.}(2013){Mori}, {Gotthelf}, {Zhang}, {An}, {Baganoff},
  {Barri{\`e}re}, {Beloborodov}, {Boggs}, {Christensen}, {Craig}, {Dufour},
  {Grefenstette}, {Hailey}, {Harrison}, {Hong}, {Kaspi}, {Kennea}, {Madsen},
  {Markwardt}, {Nynka}, {Stern}, {Tomsick}, \& {Zhang}}]{2013ApJ...770L..23M}
{Mori}, K., {Gotthelf}, E.~V., {Zhang}, S., {et~al.} 2013, \apjl, 770, L23

\bibitem[{{Most} {et~al.}(2018){Most}, {Weih}, {Rezzolla}, \&
  {Schaffner-Bielich}}]{2018PhRvL.120z1103M}
{Most}, E.~R., {Weih}, L.~R., {Rezzolla}, L., \& {Schaffner-Bielich}, J. 2018,
  Physical Review Letters, 120, 261103

\bibitem[{Olausen \& Kaspi(2014)}]{Olausen_2014}
Olausen, S.~A., \& Kaspi, V.~M. 2014, The Astrophysical Journal Supplement
  Series, 212, 6

\bibitem[{{{\"O}zel} \& {Freire}(2016)}]{2016ARA&A..54..401O}
{{\"O}zel}, F., \& {Freire}, P. 2016, \araa, 54, 401

\bibitem[{{{\"O}zel} {et~al.}(2016{\natexlab{a}}){{\"O}zel}, {Psaltis},
  {Arzoumanian}, {Morsink}, \& {Baub{\"o}ck}}]{2016ApJ...832...92O}
{{\"O}zel}, F., {Psaltis}, D., {Arzoumanian}, Z., {Morsink}, S., \&
  {Baub{\"o}ck}, M. 2016{\natexlab{a}}, \apj, 832, 92

\bibitem[{{{\"O}zel} {et~al.}(2016{\natexlab{b}}){{\"O}zel}, {Psaltis},
  {G{\"u}ver}, {Baym}, {Heinke}, \& {Guillot}}]{2016ApJ...820...28O}
{{\"O}zel}, F., {Psaltis}, D., {G{\"u}ver}, T., {et~al.} 2016{\natexlab{b}},
  \apj, 820, 28

\bibitem[{{Paschalidis} {et~al.}(2018){Paschalidis}, {Yagi},
  {Alvarez-Castillo}, {Blaschke}, \& {Sedrakian}}]{2018PhRvD..97h4038P}
{Paschalidis}, V., {Yagi}, K., {Alvarez-Castillo}, D., {Blaschke}, D.~B., \&
  {Sedrakian}, A. 2018, \prd, 97, 084038

\bibitem[{{Pavlov} {et~al.}(1996){Pavlov}, {Zavlin}, {Truemper}, \&
  {Neuhaeuser}}]{1996ApJ...472L..33P}
{Pavlov}, G.~G., {Zavlin}, V.~E., {Truemper}, J., \& {Neuhaeuser}, R. 1996,
  \apjl, 472, L33

\bibitem[{{Pons} {et~al.}(2007){Pons}, {Link}, {Miralles}, \&
  {Geppert}}]{2007PhRvL..98g1101P}
{Pons}, J.~A., {Link}, B., {Miralles}, J.~A., \& {Geppert}, U. 2007, \prl, 98,
  071101

\bibitem[{{Potekhin}(2014)}]{2014PhyU...57..735P}
{Potekhin}, A.~Y. 2014, Physics Uspekhi, 57, 735

\bibitem[{Raaijmakers {et~al.}(2019)Raaijmakers, Riley, Watts, Greif, Morsink,
  Hebeler, Schwenk, Hinderer, Nissanke, Guillot, Arzoumanian, Bogdanov,
  Chakrabarty, Gendreau, Ho, Lattimer, Ludlam, \& Wolff}]{Raaijmakers_2019}
Raaijmakers, G., Riley, T.~E., Watts, A.~L., {et~al.} 2019, The Astrophysical
  Journal, 887, L22

\bibitem[{{Ray} {et~al.}(2018){Ray}, {Arzoumanian}, {Brandt}, {Burns},
  {Chakrabarty}, {Feroci}, {Gendreau}, {Gevin}, {Hernanz}, {Jenke}, {Kenyon},
  {G{\'a}lvez}, {Maccarone}, {Okajima}, {Remillard}, {Schanne}, {Tenzer},
  {Vacchi}, {Wilson-Hodge}, {Winter}, {Zane}, {Ballantyne}, {Bozzo},
  {Brenneman}, {Cackett}, {De Rosa}, {Goldstein}, {Hartmann}, {McDonald},
  {Stevens}, {Tomsick}, {Watts}, {Wood}, \& {Zoghbi}}]{2018SPIE10699E..19R}
{Ray}, P.~S., {Arzoumanian}, Z., {Brandt}, S., {et~al.} 2018, in Society of
  Photo-Optical Instrumentation Engineers (SPIE) Conference Series, Vol. 10699,
  \procspie, 1069919

\bibitem[{{Read} {et~al.}(2009){Read}, {Lackey}, {Owen}, \&
  {Friedman}}]{2009PhRvD..79l4032R}
{Read}, J.~S., {Lackey}, B.~D., {Owen}, B.~J., \& {Friedman}, J.~L. 2009, \prd,
  79, 124032

\bibitem[{Riley {et~al.}(2019)Riley, Watts, Bogdanov, Ray, Ludlam, Guillot,
  Arzoumanian, Baker, Bilous, Chakrabarty, Gendreau, Harding, Ho, Lattimer,
  Morsink, \& Strohmayer}]{Riley_2019}
Riley, T.~E., Watts, A.~L., Bogdanov, S., {et~al.} 2019, The Astrophysical
  Journal, 887, L21

\bibitem[{{Ruderman} \& {Sutherland}(1975)}]{1975ApJ...196...51R}
{Ruderman}, M.~A., \& {Sutherland}, P.~G. 1975, \apj, 196, 51

\bibitem[{{Shibanov} {et~al.}(1992){Shibanov}, {Zavlin}, {Pavlov}, \&
  {Ventura}}]{1992A&A...266..313S}
{Shibanov}, I.~A., {Zavlin}, V.~E., {Pavlov}, G.~G., \& {Ventura}, J. 1992,
  \aap, 266, 313

\bibitem[{{Sieniawska} {et~al.}(2018){Sieniawska}, {Bejger}, \&
  {Haskell}}]{2018A&A...616A.105S}
{Sieniawska}, M., {Bejger}, M., \& {Haskell}, B. 2018, \aap, 616, A105

\bibitem[{{Sieniawska} {et~al.}(2019){Sieniawska}, {Turcza{\'n}ski}, {Bejger},
  \& {Zdunik}}]{2019A&A...622A.174S}
{Sieniawska}, M., {Turcza{\'n}ski}, W., {Bejger}, M., \& {Zdunik}, J.~L. 2019,
  \aap, 622, A174

\bibitem[{{Suleimanov} {et~al.}(2009){Suleimanov}, {Potekhin}, \&
  {Werner}}]{2009A&A...500..891S}
{Suleimanov}, V., {Potekhin}, A.~Y., \& {Werner}, K. 2009, \aap, 500, 891

\bibitem[{{Turolla} \& {Nobili}(2013)}]{2013ApJ...768..147T}
{Turolla}, R., \& {Nobili}, L. 2013, \apj, 768, 147

\bibitem[{{Watts}(2019)}]{2019arXiv190407012W}
{Watts}, A.~L. 2019, arXiv e-prints, arXiv:1904.07012

\bibitem[{{Zhang} \& {\it et al.}(2019)}]{2019SCPMA..6229502Z}
{Zhang}, S., \& {\it et al.} 2019, Science China Physics, Mechanics, and
  Astronomy, 62, 29502

\end{thebibliography}

\end{document}